\documentclass[pof, reprint, longbibliography, nofootinbib]{revtex4-1}

\usepackage{soul}          
\usepackage{float}
\usepackage{amssymb}
\usepackage{amsmath}
\usepackage[dvipsnames]{xcolor}
\usepackage[colorlinks=true]{hyperref}
\usepackage[capitalise]{cleveref}
\usepackage{etoolbox}
\usepackage{subfigure}
\usepackage{fancyhdr}

\usepackage{todonotes}
\usepackage[utf8]{inputenc}
\usepackage{multirow}
\usepackage[pagewise]{lineno}
\usepackage[normalem]{ulem}
\usepackage{color}

\newcommand{\red}[1]{{\color{red}#1}}

\begin{document}




\title{Deep Neural Networks for Nonlinear Model Order Reduction of Unsteady Flows}

\author{Hamidreza Eivazi$^{1}$}
\email[]{hamid.eivazi@ut.ac.ir}

\author{Hadi Veisi$^{1}$}
\email[Corresponding author; ]{h.veisi@ut.ac.ir}

\author{Mohammad Hossein Naderi$^{1}$}
\email[]{mhnaderi1994@ut.ac.ir}

\author{Vahid Esfahanian$^{2}$}
\email[]{evahid@ut.ac.ir}

\affiliation{$^{1}$Faculty of New Sciences and Technologies, University of Tehran, Tehran 1439957131, Iran}
\affiliation{$^{2}$School of Mechanical Engineering, College of Engineering, University of Tehran, Tehran 14395515, Iran}


\begin{abstract}
    Unsteady fluid systems are nonlinear high-dimensional dynamical systems that may exhibit multiple complex phenomena both in time and space. Reduced Order Modeling (ROM) of fluid flows has been an active research topic in the recent decade with the primary goal to decompose complex flows to a set of features most important for future state prediction and control, typically using a dimensionality reduction technique. In this work, a novel data-driven technique based on the power of deep neural networks for reduced order modeling of the unsteady fluid flows is introduced. An autoencoder network is used for nonlinear dimension reduction and feature extraction as an alternative for singular value decomposition (SVD). Then, the extracted features are used as an input for long short-term memory network (LSTM) to predict the velocity field at future time instances. The proposed autoencoder-LSTM method is compared with non-intrusive reduced order models based on dynamic mode decomposition (DMD) and proper orthogonal decomposition (POD). Moreover, an autoencoder-DMD algorithm is introduced for reduced order modeling, which uses the autoencoder network for dimensionality reduction rather than SVD rank truncation. Results show that the autoencoder-LSTM method is considerably capable of predicting fluid flow evolution, where higher values for coefficient of determination $R^{2}$ are obtained using autoencoder-LSTM compared to other models.

\end{abstract}


\maketitle 
\let\thefootnote\relax\footnotetext{This article may be downloaded for personal use only. Any other use requires prior permission of the author and AIP Publishing. This article appeared in: Phys. Fluids \textbf{32}, 105104 (2020), and may be found at: \url{https://doi.org/10.1063/5.0020526}.}

\section{Introduction}

Fluid flow around aerodynamic configurations can experience complex nonlinearities with a wide range of spatial and temporal features. These are mainly related to nonlinear convection term and turbulence that often cannot be treated by simple linearization. Many of complex fluid flows evolve on a low-dimensional subspace that may be characterized by dominant spatiotemporal coherent structures. It is of interest in the analysis of unsteady fluid flows to extract dominant features and introduce a reduced model of the complex system based on physically important features. This performs typically through the modal decomposition of a numerical or experimental dataset of the flow-field.  However, a problem arises when attempting model reduction of unsteady flows, where long term transient phenomena need to be predicted accurately. Some examples of common unsteady flow features and phenomena are von Kármán shedding, Kelvin–Helmholtz instability, and vortex pairing/merging \cite{Taira2017}. Reduced order modeling of the fluid systems is a challenging task due to the existence of these complex phenomena at multiple spatiotemporal scales. It means that they are difficult to reduce to a low-dimensional subspace without losing at least some of these scales. During the last three decades, several efforts in theoretical foundations, numerical investigations, and methodological improvements have made it possible to develop general ideas in reduced order modeling and to tackle several problems arising in fluid dynamics \cite{Stankiewicz2016,Mohebujjaman2019,Star2019}. Proper orthogonal decomposition (POD) \cite{Lumley1967,Lumey1970}, dynamic mode decomposition (DMD) \cite{Schmid2010}, and Koopman analysis \cite{Rowley2009} are some of the well-known reduced order methods (ROMs) in the field of fluid dynamics. POD was first introduced by Lumley \cite{Lumley1967} to the fluid dynamics community for the study of the turbulent boundary layer. It is a modal decomposition method designed to identify the features (a set of orthogonal modes) of flow most important for reconstructing a dataset. POD modes are not necessarily optimal for modeling dynamical systems while the modes do not depend on the time evolution/dynamics encoded in the data \cite{Tu}. DMD is a method for analyzing the time evolution of a dynamical system. It is originated in the fluid dynamics community and first was introduced by Schmid and Sesterhenn \cite{Schmid2008}. Only a year later, Rowley et al.~\cite{Rowley2009} presented a technique for describing the global behavior of complex nonlinear flows by decomposing the flow into modes determined from spectral analysis of the Koopman operator \cite{Koopman1931}. They showed that DMD could be considered as an algorithm that calculates an approximate Koopman decomposition. Schmid~\cite{Schmid2010} followed this with his article in 2010 and demonstrated the DMD capability in providing an accurate decomposition of complex systems into coherent spatiotemporal structures that may be used for short-time future state prediction and control. DMD is an equation-free data-driven method based on the power of the singular value decomposition (SVD), and it has been employed for modal analysis of a variety of fluid flows \cite{Alekseev2016,Bistrian2017}.

Deep learning is a subset of machine learning methods based on artificial neural networks, which is capable of extracting hidden information with multiple levels of representation from nonlinear and complex dynamical systems \cite{LeCun2015}. Lately, deep learning has made its mark in various areas such as virtual assistants, image and video processing, speech recognition, genetics and disease diagnosis \cite{Sainath2013,Krizhevsky2012,Abdolhosseini2019}.

In recent years, efficient strategies have emerged in the field of fluid mechanics based on machine learning (ML) and deep neural networks (DNNs). Complementary information on the use of ML for fluid dynamics can be found in many recent reviews~\cite{willard_integrating_2020, rabault_deep_2020, pandey_perspective_2020, fukami_assessment_2020, Brunton2020, Kutz2017}. It was shown that the autoencoder network could be efficiently used as a nonlinear principal component analysis (PCA) technique for dimensionality reduction of the fluid flows rather than the linear SVD. Moreover, the ability of recurrent neural networks (RNNs) for prediction of the sequential data brings the idea of using RNNs for learning dynamics and future state estimation of the dynamical systems. The advantage of using deep neural networks rather than the Galerkin projection technique has been stated in several research studies \cite{Pawar2019}. Besides, a type of RNNs, the long short-term memory (LSTM), which has shown attractive potential in the modeling of the sequential problems such as speech modeling and language translation, is used for prediction of the flow evolution, e.g., learning dynamic and prediction of the turbulent shear flows \cite{Schlatter2019}. Yang et al.~\cite{yang_improving_2020}, used field inversion and ML to enhance the four-equation $k–\omega–\gamma–A_{r}$ transition model. Moreover, ML strategies have been used for extracting information from high-fidelity data to inform low-fidelity numerical simulations for higher accuracy. The specific aim is to use DNNs to build an improved representation of the Reynolds stress anisotropy tensor from high-fidelity simulation data \cite{ling2016,Duraisamy2019}. Besides, deep neural networks have been used for learning the physics of the flow over aerodynamic structures to hasten the process of geometrical optimization \cite{Savonius2019}. Recently, deep reinforcement learning has been used for active flow control in many studies~\cite{rabault_artificial_2019, tang_robust_2020, beintema_controlling_2020}. In brief, dimensionality reduction, feature extraction, super-resolution, reduced order modeling, turbulence closure, shape optimization, and flow control are some of the critical tasks in fluid dynamics that could be enhanced with the implementation of the ML algorithms.

Unsteady flows are, in essence, high dimensional dynamical systems. The use of ML for model-free prediction of spatiotemporal dynamical systems purely from the system's past evolution is of interest to both scientists and industries in the field of fluid dynamics. Therefore, the development of non-intrusive reduced order models (NIROMs) is of particular interest in the fluid dynamics community. In the context of ROM, both dimensionality reduction and future state prediction could be advanced from deep learning strategies. Baldi and Hornik \cite{Baldi1989} showed that an autoencoder network with one hidden layer and linear activation functions could closely resemble the standard POD/PCA decomposition. Wang et al.~\cite{Wang2016} used a deep autoencoder architecture for reduced order modeling and dimensionality reduction of distributed parameter systems. Liu et al.~\cite{Liu2019} proposed a novel compression method based on generative adversarial network (GAN), which had significant advantages in compression time. More recently, Murata et al.~\cite{Murata2020} developed a new nonlinear mode decomposition method based on a convolutional neural network (CNN) autoencoder. Implementation of nonlinear activation functions leads to lower reconstruction errors than POD, and nonlinear embedding considered the key to improving the capability of the model. Although the deep learning method is used to make dimensionality reduction, they do not take advantage of its predictive capability. Moreover, Guastoni et al.~used convolutional-network models for the prediction of wall-bounded turbulence from wall quantities \cite{Guastoni2020a,Guastoni2020b}.

RNNs are of particular interest to fluid mechanics due to their ability in learning and prediction of the sequential data. The renewed interest in RNNs has largely attributed to the development of the LSTM algorithms. Commonly, the Galerkin projection has been used to obtain the evolution equations for the lower order system with POD. However, POD-Galerkin method is an intrusive ROM (IROM), and several studies have been conducted using ANNs \cite{San2019,San2018}, RNNs \cite{Pawar2019}, and LSTM networks \cite{Wang2018,Mohan2018,ahmed2019} to construct an NIROM based on POD modes. Li et al.~\cite{li_deep_2019} investigated a new model to improve ROM's predicting ability for varying operating conditions. They used the LSTM network for large training datasets and sampling space and assessed the performance of the method by a NACA 64A010 airfoil pitching and plunging in the transonic flow condition over various Mach numbers. Renganathan et al.~\cite{renganathan_machine_2020} proposed an ML method to build ROMs with LSTM and POD. They showed that this procedure preserves accuracy with a significantly more inexpensive computational cost. Their ML method is physics-informed and restrained by the utilization of an interpretable encoding. Halder et al.~\cite{halder_deep_2020} presented a method for modeling transonic airfoil–gust interaction, which included two fundamental steps; A dimensionality reduction method where the Discrete Empirical Interpolation was employed, and the compressed state was trained utilizing the LSTM for the reconstruction of the flow field. Maulik et al.~\cite{maulikNonautoregressiveTimeseriesMethods2020} formed a non-autoregressive time series method in order to forecast linear reduced-basis time records of forwarding bases. They showed that non-autoregressive counterparts of sequential learning techniques (e.g., LSTM) significantly increase machine-learned ROMs' stability and assessed their method on the shallow water equations. Tatar and Sabour~\cite{tatarReducedorderModelingDynamic2020} introduced a nonlinear ROM of dynamic stall employing a fuzzy inference system (FIS) and the adaptive network-based FIS (ANFIS). They used the Gram–Schmidt orthogonalization method to build a higher dimension of the input variables. The outcomes reveal that great modeling is obtained by ANFIS within the new structure of the inputs and the corresponding aerodynamic coefficients utilizing ten percent of data. Pawar et al.~\cite{pawarDatadrivenRecoveryHidden2020} employed POD as a dimensionality reduction technique to build orthonormal bases and a Galerkin projection to create a dynamical core of a system. The Galerkin projection also adjusted at each time-step by employing an LSTM network to include hidden physics. Furthermore, a Grassmann manifold method was selected to interpolate basis functions to unnoticed parametric restrictions.

Besides, data-driven finite dimensional approximations of the Koopman operator has been received significant attention in recent years, in particular, for problems dealing with complex spatiotemporal behaviors such as unsteady fluid flows. DMD, in its original formulation, implicitly utilizes linear observables of the state of the system. Extended DMD \cite{Williams2015} and Hankel-DMD \cite{Arbabi2017} were proposed to include a richer set of observables that spans a Koopman invariant subspace. Recently, several works have been done, introducing fully data-driven approaches for learning Koopman embedding using deep neural networks (DNNs) \citep{Lusch2018,Takeishi2017,Li2017}.  Morton et al.~\cite{Morton2018} presented a new architecture based on the work by Takeishi et al.~\cite{Takeishi2017} to predict the time evolution of a cylinder system. Their approach was grounded in Koopman's theory and aimed to estimate a set of functions that transform data into a form in which a linear least-squares regression fits well. However, it has been shown by Khodkar et al.~\cite{khodkar2019} that the linear combination of a finite number of modes may not accurately reproduce the nonlinear characteristics of a chaotic dynamics for a reasonably long period of time. They showed that adding nonlinearities to the linear model as a forcing term leads to an excellent short-term prediction of several well-known prototypes of chaos. Moreover, Eivazi et al.~\cite{Eivazi2020} showed that the Koopman framework with nonlinear forcing (KNF) provides accurate predictions of the dynamical evolution of the coefficients of a low-order model for near-wall turbulence \cite{moehlis_et_al}.

To this end, it has been shown that adding nonlinearity to the dimensionality reduction process by using nonlinear activation functions in the autoencoder architecture can effectively enhance the efficiency of the dimensionality reduction process. However, the majority of research studies have been dedicated to the development of reduced order models based on linear compression techniques such as SVD. Moreover, promising capability of the LSTM network in the learning of the complex systems dynamic brings the opportunity to implement LSTM network for future time prediction rather than a linear mapping of the previous observations to the future time instances using the DMD algorithm. In this regard, this paper presents a novel reduced order model based on deep neural networks. A deep autoencoder network with nonlinear activation functions is designed and implemented for nonlinear dimensionality reduction and dominant features extraction, and then, LSTM network is used for prediction of the flow evolution. A sequence of the extracted features from the autoencoder network is the input of the LSTM, and the output is the flow-field in the future time step. Train and test sets are acquired from numerical simulation of the flow around a cylinder and an oscillating airfoil. Two test cases are examined for the cylinder. One at a constant Reynolds number $Re$ of 3900 inducing single frequency vortex shedding, and the other one at gradually decreasing Reynolds number from 3355 to 676, which leads to decay of the vortex shedding behind the cylinder, and consequently, a variable frequency dynamical system. The airfoil oscillates sinusoidally at $Re$ number of $1.35 \times 10^{5}$. The performance of the proposed autoencoder-LSTM method in future state prediction of the flow is compared with the DMD and POD based models. Moreover, an autoencoder-DMD algorithm is introduced for reduced order modeling, which uses the autoencoder network for dimensionality reduction rather than SVD rank truncation. The main novelties of this research study are listed below:
\begin{itemize}
\item Introduce a novel data-driven method based on deep neural networks for nonlinear reduced order modeling of complex unsteady fluid flows
\item Implementation of autoencoder network for nonlinear dimensionality reduction and feature extraction of the fluid systems
\item Future state estimation of a nonlinear dynamical system using LSTM network from its features extracted by the autoencoder network
\item Comparison of linear and nonlinear non-intrusive ROM frameworks 
\end{itemize}

\section{Numerical Analysis and Data Sets}

\begin{figure}[hbtp]
    \centering
    \includegraphics[width=0.8\columnwidth]{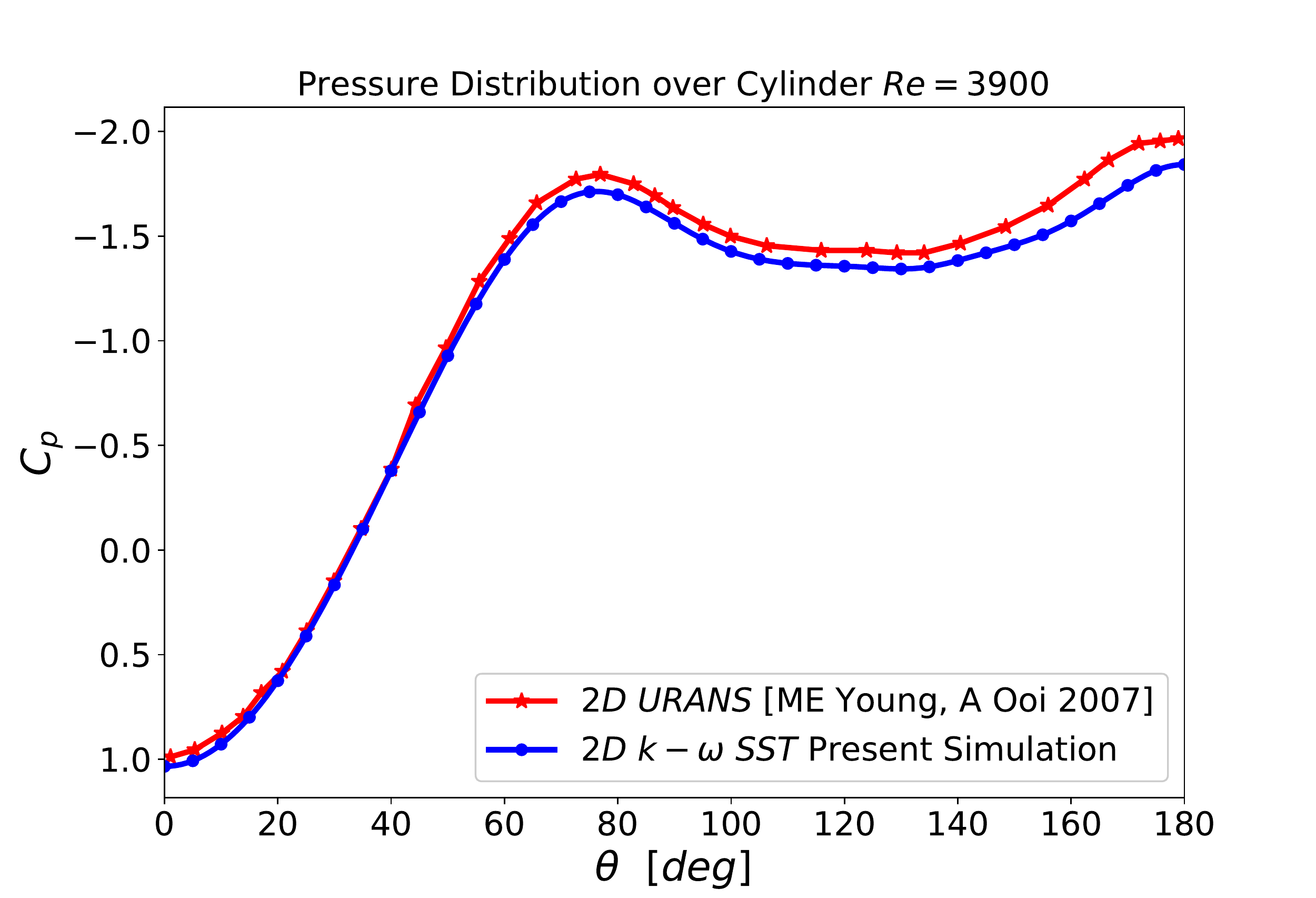}
    \caption{Pressure distribution over cylinder, comparison of present simulation with URANS simulation of Young and Ooi \cite{Young2007}}
    \label{validation}
\end{figure}

Two-dimensional Navier-Stokes equations are solved using finite volume method (FVM). Temporal and spatial properties are discretized with the second-order implicit scheme and second-order upwind scheme, respectively. The computational domain is a circle with O-type mesh and velocity inlet and pressure outlet as the boundary conditions. $k-\omega~SST$ turbulence model is adopted and $y^{+}$ value is checked to be lower than one. 
For the unsteady flow over the cylinder, two test cases are examined. First, the flow around a cylinder at Reynolds number $Re = U_{\infty} D / \nu $ of 3900 (cylinder test 1) is simulated where $\nu$ is the kinematic viscosity, $D$ is the cylinder diameter, and $U_{\infty}$ is the free-stream velocity. The computational domain is a circle with a radius of 20D. The O-type grid is generated with 360 equally sized grid point in the azimuthal direction, but with grid stretching in the radial direction with the total cell number of 98878. \Cref{validation} represents the pressure distribution over the cylinder for the present simulation against the URANS simulation of \cite{Young2007}, which is shown a good agreement between the results. 
Simulation is conducted with a time step of 0.2, which results in 357 time steps per each cycle period. Snapshots are extracted from every seven time steps; consequently, 51 snapshots for each cycle. Calculations are performed for 20 complete cycles; four cycles are considered as a training set, and the next four cycles are picked for testing the networks. 

\begin{figure}[hbtp]
    \centering
    \includegraphics[width=0.8\columnwidth]{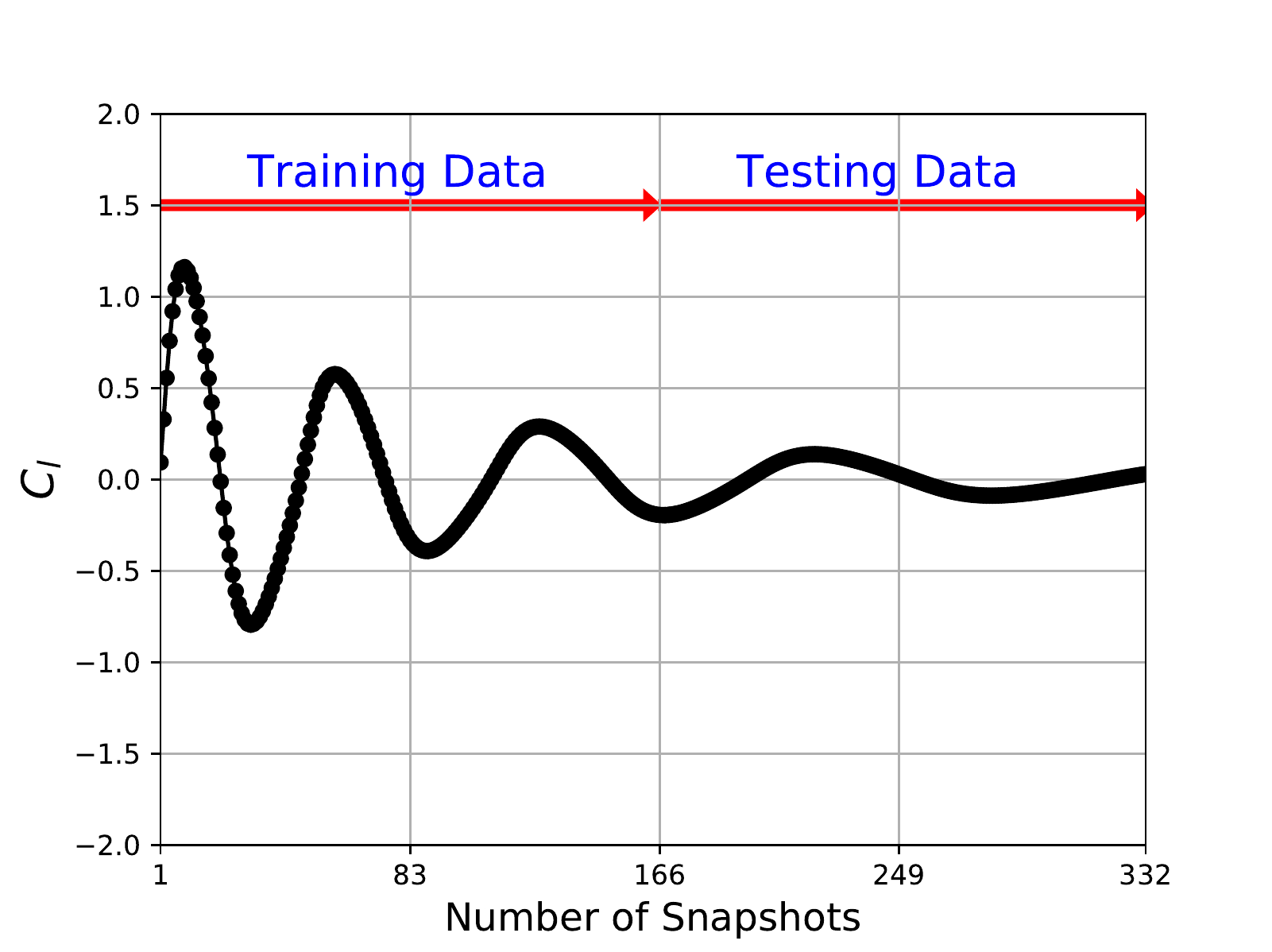}
    \caption{Variation of Lift coefficient with time for the second test case}
    \label{cl2}
\end{figure}

\begin{figure}[hbtp]
    \centering
    \includegraphics[width=0.8\columnwidth]{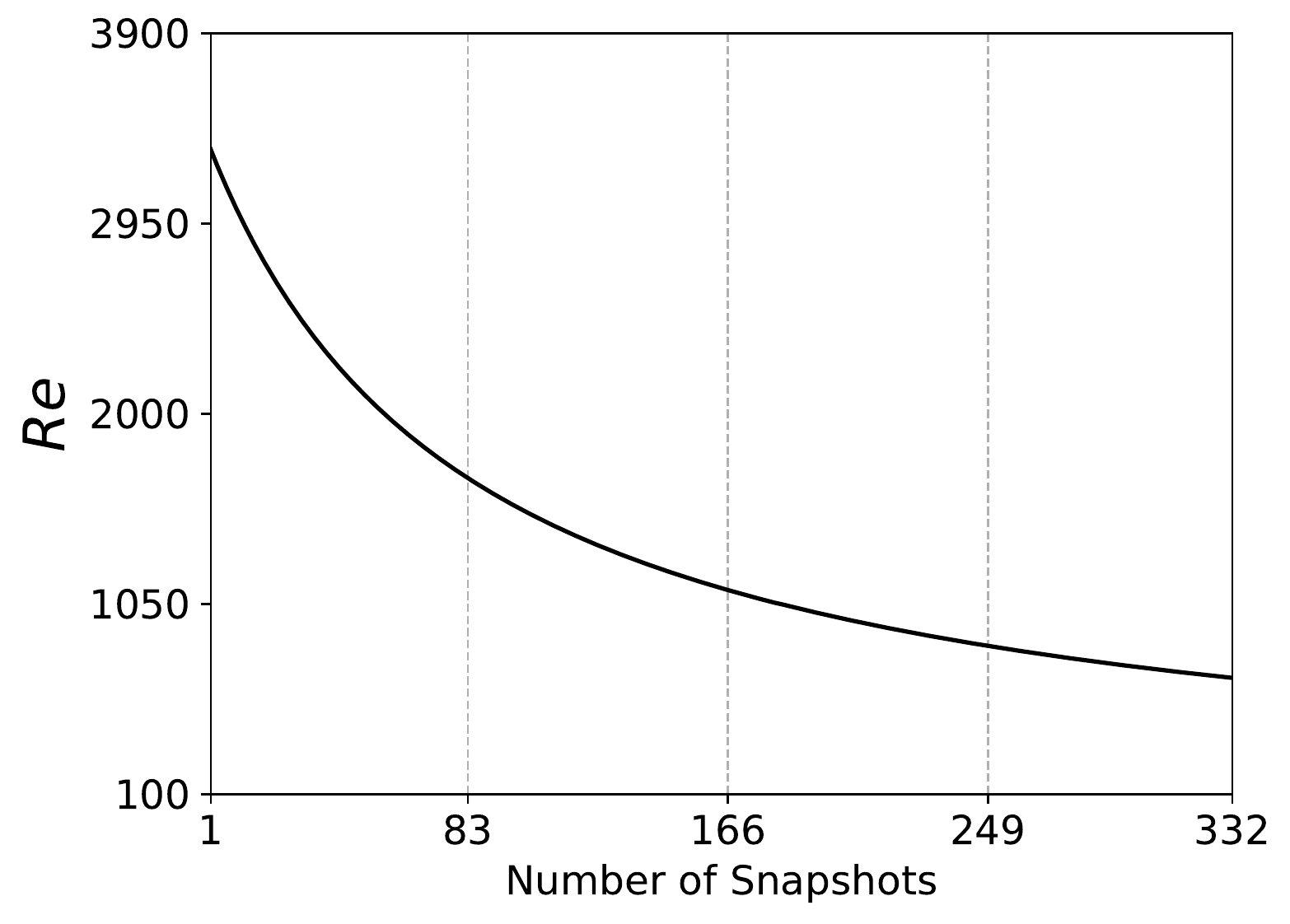}
    \caption{Reynolds number at the collected snapshots for the second test case}
    \label{Re}
\end{figure}

At the second test case for the cylinder, free-stream velocity $U_{\infty}$ is decreased relative to the time from its value corresponding to $Re = 3900$ as $U_{\infty} = U_{\infty, Re = 3900}/t$, where $t$ is the time. The simulation time step is 0.2, and snapshots are collected from the velocity field at every ten time steps. 332 snapshots are extracted from the wake of the cylinder, and 166 snapshots are used as the training set. Variation of lift coefficient besides training and testing data sets are shown in \cref{cl2}. The gradual decrease of the free-stream velocity weakens the vortex shedding and leads to the change in its frequency. Reynolds number varies from 3355 to 676, as it is depicted in \cref{Re}.

\begin{figure}[hbtp]
    \centering
    \includegraphics[width=0.8\columnwidth]{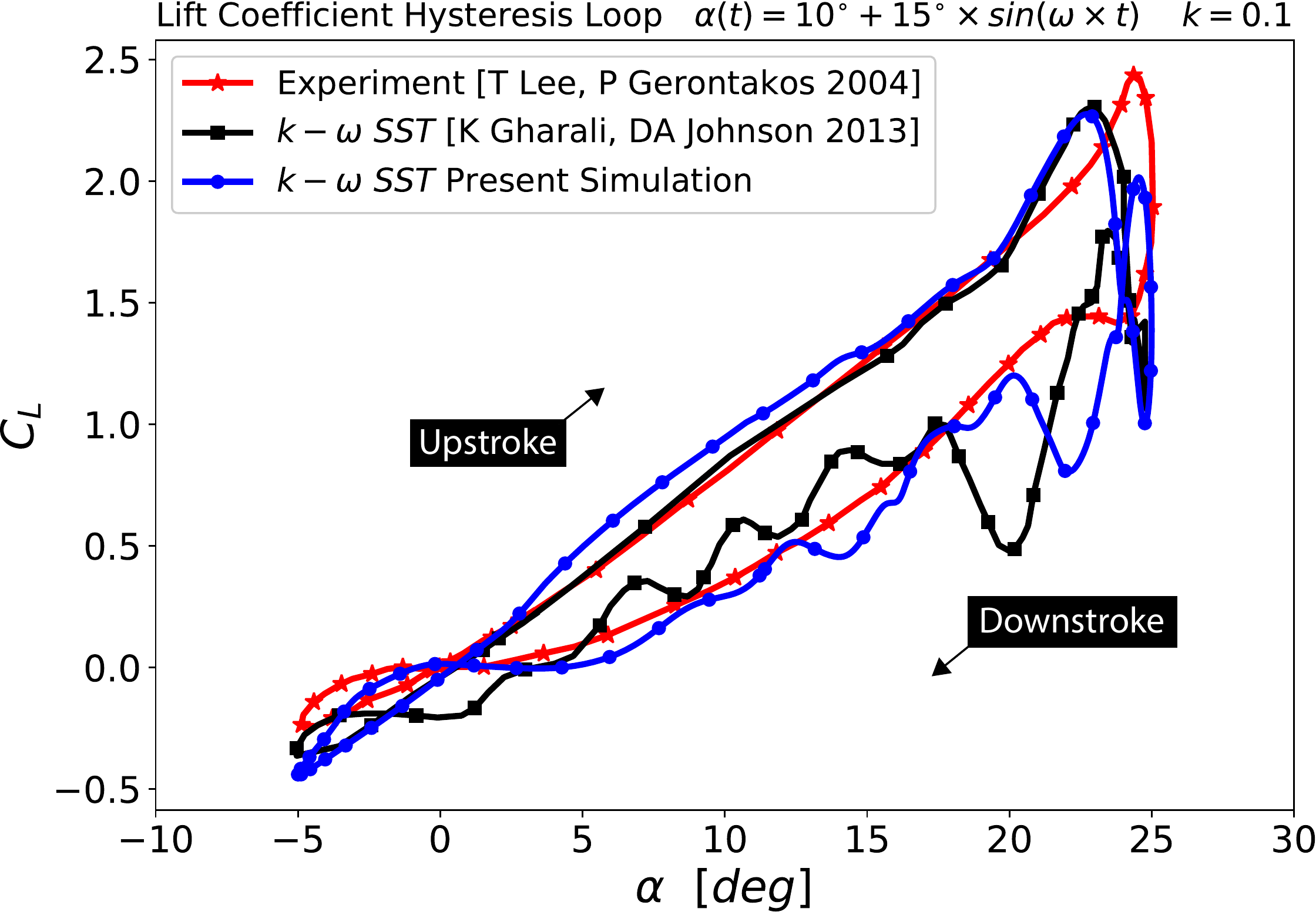}
    \caption{Lift coefficient versus angle of attack for the pitch oscillating airfoil; comparison of the present simulation and experimental \cite{Lee} and numerical \cite{Gharali2013} data}
    \label{airfoilvalid}
\end{figure}

For the oscillating airfoil test case, the flow around a pitch oscillating NACA0012 airfoil is simulated at Reynolds number $Re=U_{\infty} C/ \nu$ of $1.35\times10^{5}$, where $C$ is the airfoil chord. Airfoil oscillates sinusoidally $(\alpha=\alpha_{mean}+\alpha_{amp}\times\sin(\Omega t))$ about its $\frac{1}{4}$ chord with the reduced frequency $k=\Omega C/2 U_{\infty}$ of 0.1 according to the \cite{Lee} test case. $\alpha_{mean}$ and $\alpha_{amp}$ represent mean angle of attack and amplitude of oscillations, respectively. Airfoil oscillation is modeled with the sliding mesh technique. For using dimensional reduction methods on moving meshes, the data collected at each time step from grid points can be interpolated to a single stationary grid and then snapshots can be constructed from these fixed points.
The data on the dynamic mesh is interpolated on a constant mesh using K-nearest neighbors (KNN) machine learning algorithm. The principle behind the KNN method is to detect a predefined number of training instances that are close to the new point and compute the value based on the mean of these nearest neighbors' values. In this method, the velocity of points which are inside the solid structure (e.g., airfoil) at each time step is set to zero.
Complementary information about data interpolation using ML techniques can be found in work by Naderi et al.~\cite{Naderi2019}.~\Cref{airfoilvalid} presents the lift coefficient versus angle of attack obtained from present simulation against experimental and numerical data \cite{Lee,Gharali2013}. For this test case, the data corresponding to two cycles of airfoil's oscillations are used as the training data set and the next two cycles are predicted.

\section{Methodology}

To extract dominant features of the unsteady flow-field around the test cases, an autoencoder network is designed and trained with the snapshots obtained from CFD simulation. These features are organized as temporal sequences that are used as the inputs of the LSTM network with the aim of the prediction of the flow-field at future time steps. In contrast with the SVD, which is a basic part of modal decomposition techniques, autoencoder with nonlinear activation functions provides a nonlinear dimensionality reduction. In the following, an elaboration on the networks architecture, inputs and outputs of each network, training and optimization algorithms, and selection of hyperparameters is presented.
  
\subsection{Autoencoder Neural Network}
\label{s-auto}

\begin{figure}[hbtp]
    \centering
    \includegraphics[width=0.9\columnwidth]{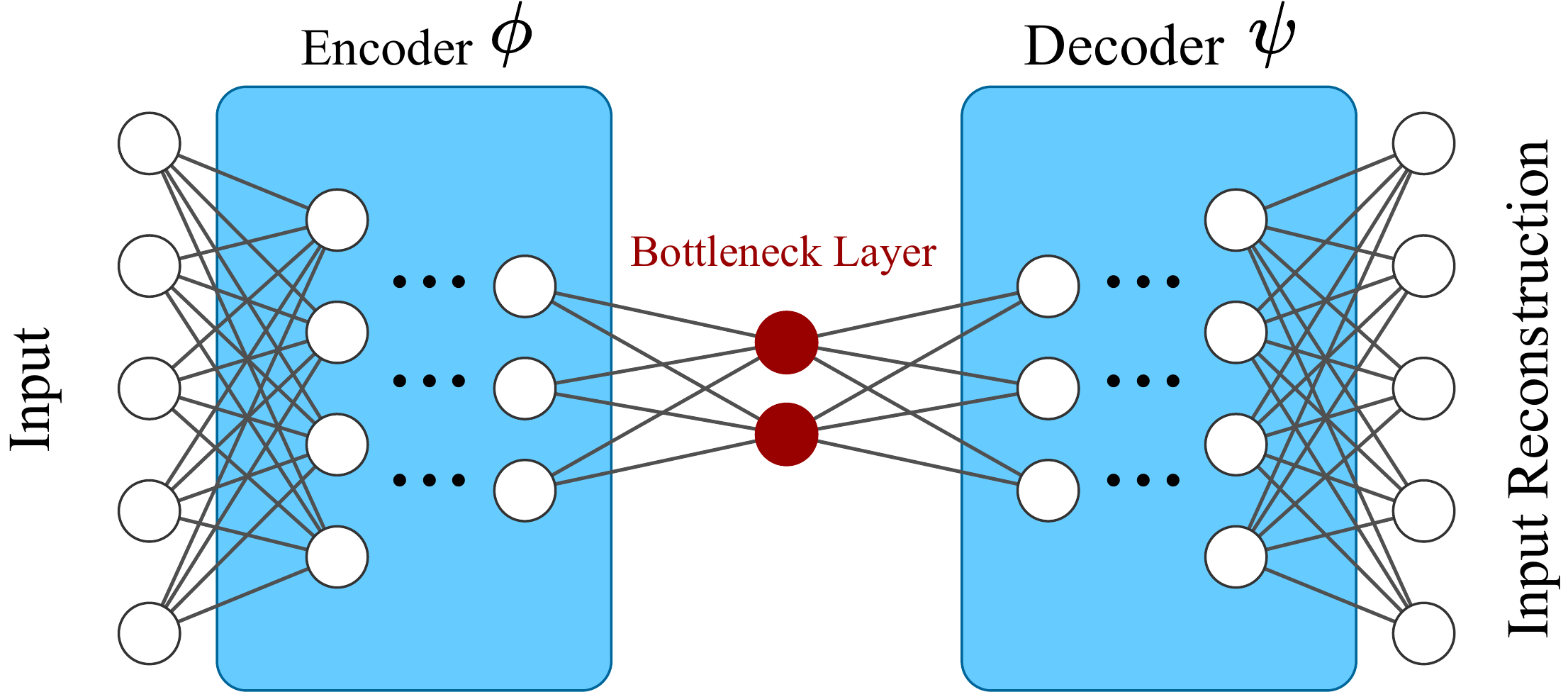}
    \caption{The architecture of autoencoder network}
    \label{auto}
\end{figure}

Autoencoder neural networks are a type of network architecture developed for unsupervised feature extraction and dimensionality reduction. The main incentive for utilizing autoencoder architecture in the context of ROM is to perform a nonlinear dimensionality reduction to possibly enhance the efficiency of the process. In this architecture, the inputs and the outputs are the same with a size larger than the dimension of the hidden layers. The network comprises of two parts, namely, encoder and decoder. The encoder part, $\phi$, maps the inputs to a latent space, $g$, with a lower dimension by decrease of the number of neurons in hidden layers. Then, the decoder part, $\psi$, projects the latent space back to the original space according to \cref{Eqauto}. The middle layer between encoder and decoder is called the bottleneck layer and may be used for the extraction of the most dominant features from the input data. In this study, the input and output data of the autoencoder network are the velocity magnitudes $V$ on the computational grid points obtained from CFD simulation of the flow-field over the test cases. The data on the two-dimensional CFD grid is first flattened and then fed to the network. The architecture of the autoencoder network, encoder and decoder parts, and the bottleneck layer are depicted in \cref{auto}.

\begin{equation}
    g(t) = \phi(V(t)),~~~~ \tilde{V}_{rec}(t)= \psi(g(t))
    \label{Eqauto}
\end{equation}

\subsection{Long Short-Term Memory Network}
\label{s-lstm}

\begin{figure}[!htbp]
   \centering
       \includegraphics[width=0.8\columnwidth]{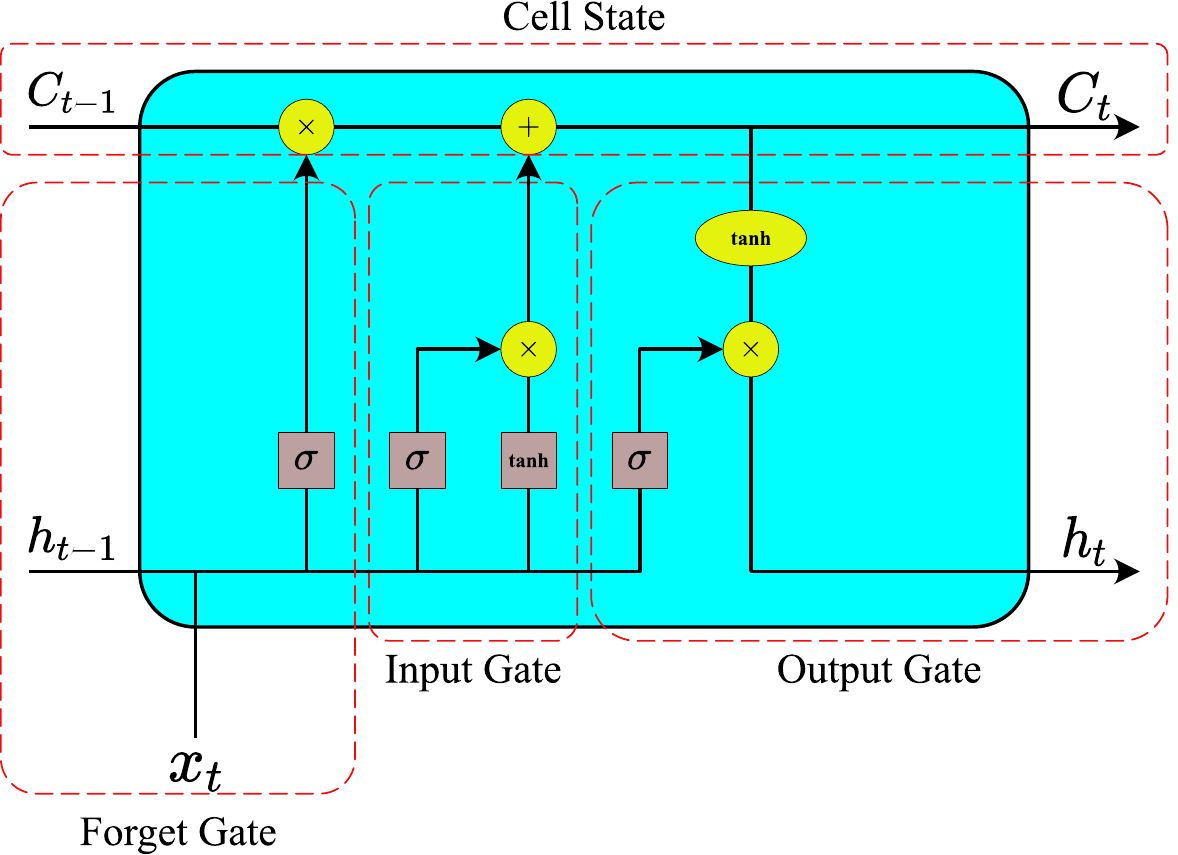}
   \caption{Schematic of the LSTM cell}
   \label{fig:LSTM_Cell}
\end{figure}

Long short-term memory (LSTM) was proposed by Hochreiter and Schmidhuber in 1997 \cite{LSTM}. LSTM is an artificial recurrent neural network (RNN) architecture designed for learning of the sequential data such as numerical or experimental time series and to overcome the forgetting problem in conventional RNNs \cite{Goodfellow2016}. An LSTM neural network unit is comprised of an input gate, a cell state, a forget gate, and an output gate (\cref{fig:LSTM_Cell}). The cell state can transfer important data throughout the processing of the sequence. Therefore the information from the earlier sequences or time steps can move to later time steps, decreasing the impacts of short-term memory. While the cell state tries to transfer the information, data get appended or eliminated to the cell state by gates. The gates are various neural networks that choose which data is permitted on the cell state.

\begin{itemize}
    \item \textbf{Forget Gate}
    
    Data from the previous time step ($h_{t-1}$) and information from the current time step ($x_{t}$) is given through the sigmoid function. Values come out within 0 and 1. If the output is over some trained parameter, then the cell state is entirely reset to 0, mostly forgetting its long term memory.
    
    \item \textbf{Input Gate}
    
    The input gate decides if the data is required for the long term. Firstly, we transfer the output of the hidden state ($h_{t-1}$) and the labeled data input ($x_{t}$) into a sigmoid function. That determines which values will be updated by changing the amounts to be between 0 and 1: 0 indicates not essential, and 1 means important. The hidden state and current input also are passed into the $\tanh$ function to assist in regulating the neural network. Next, the $\tanh$ output is multiplied with the sigmoid function output. The sigmoid function output will select which data is critical to hold from the $\tanh$ output.

    \item \textbf{Cell State}
    
    The $C_{t}$ indicates the current cell state of the LSTM neural network. The cell state is an array of numbers that are transferred through all cells in its way. The cell state is multiplied with the forget array. This has the chance of declining values in the cell state if it gets multiplied by numbers close to zero. Next, the output from the input gate and is added, which updates the cell state to a new state. That provides the latest cell state.

    \item \textbf{Output Gate}
    
    The output gate provides the following hidden state array for the next cell ($h_{t}$). The former hidden state and the current input is passed toward a sigmoid function. Next, the recently changed cell state is transferred to the $\tanh$ function. The $\tanh$ output is multiplied by the sigmoid output to determine what data the hidden state should take. This is the short-term memory character of the LSTM neural network.
\end{itemize}

\subsection{Autoencoder-LSTM method for ROM}
\label{method_autoLSTM}
\begin{figure*}[hbtp]
    \centering
    \includegraphics[width=0.7\textwidth]{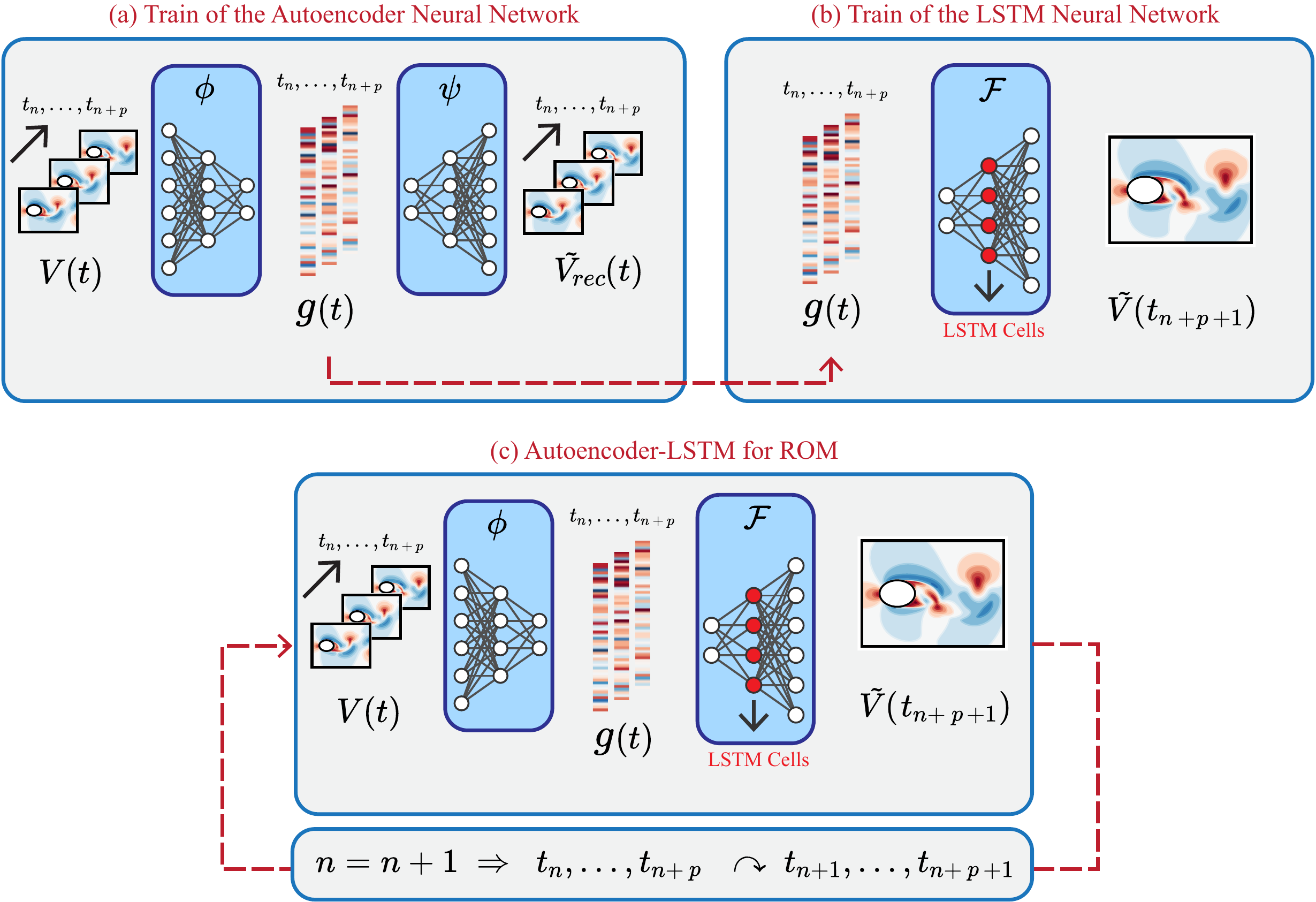}
    \caption{Autoencoder-LSTM reduced order modeling framework; a) train of the autoencoder network, given snapshot data $V(t)$ return unsteady flow features $g(t)$ and reconstructed data $\tilde{V}_{rec}(t)$; b) train of the LSTM network, given a sequence of extracted flow features of $g(t)$ compute the data for the next time step $\tilde{V}(t_{n+p+1})$; c) autoencoder-LSTM for ROM, given a sequence of snapshots $V(t)$ return unsteady flow features $g(t)$ and predict flow-field for the next time step of the sequence $\tilde{V}(t_{n+p+1})$ and iterate to predict the unsteady flow over time.}
    \label{flowchart_AutoLSTM}
\end{figure*}

In this paper, the power of autoencoder and LSTM networks in nonlinear dimension reduction and learning of the sequential data is leveraged for non-intrusive reduced order modeling of unsteady flows. The schematic of the proposed method is presented in \cref{flowchart_AutoLSTM}. The first step is to train the autoencoder network using the snapshot data collected from the flow-field with the aim of nonlinear dimension reduction and feature extraction (\cref{flowchart_AutoLSTM}\red{a}). The data on the two-dimensional grid is first flattened and then fed to the network. Each time step data $V(t)$, which is a vector of state variables at time $t$, is mapped to the latent space through the function $\phi$ as $g(t)$ with much lower dimension,  $n_{g} \ll n_{input}$. Then, $g(t)$ is projected back to the original space to reconstruct the input data, $\tilde{V}_{rec}(t)$. The data on the latent space, $g(t)$, is then used as the input to train the LSTM network with the aim of future prediction (\cref{flowchart_AutoLSTM}\red{b}). The input of the LSTM network is a sequence of $g$ as $[g(t_{n}), ... , g(t_{n+p})]$ with the length of $p$ and the output is the flow-field data at the next time step of the input sequence $\tilde{V}(t_{n+p+1})$. Note that in this way, $p$ previous time units are used for prediction of the next time step. For testing and using the autoencoder-LSTM method for reduced order modeling, an iterative procedure is conducted. An initial sequence is first mapped to the latent space using the encoder ($\phi$) part of the autoencoder. The data on the latent space is fed to the LSTM network to predict the flow-field data for the next time step of the input sequence. The predicted value is then stacked to the input sequence ignoring the first snapshot data to prepare a sequence one step further in time. The new sequence is fed again to the network to predict the next time step, and the procedure is performed iteratively for future prediction (\cref{flowchart_AutoLSTM}\red{c}). All of the neural network models are created using a machine learning software framework developed by Google Research called TensorFlow \cite{TensorFlow}. The autoencoder and LSTM networks are trained by the feedforward and backpropagation algorithms. For all the training runs in this work, a variant of the stochastic gradient descent algorithm called adaptive moment estimation (Adam) \cite{Adam} is utilized with the learning rate of 0.001. Adam has an adaptive learning rate method, which is commonly used to train deep networks. As a standard choice for regression problems, the mean squared error (MSE) is used as the loss function according to \cref{MSE}. For all the training, the batch size is equal to 32.

Two different metrics are used for error estimation. The first metric is Coefficient of Determination ($R^{2}$) which is defined in \cref{R2}. $R^{2}$ is commonly between zero and one.
\begin{equation}
    \label{R2}
    R^{2}=1-\frac{\sum_{i=1}^{n}\left(y_{i}-\widehat{y_{i}} \right) ^{2}}{\sum_{i=1}^{n}\left(y_{i}-\overline{y} \right) ^{2}}
\end{equation} 
The second metric for error calculation is MSE, which is also considered as the loss function for training of the networks (\cref{MSE}).
\begin{equation}
    \label{MSE}
   \text{MSE} =\frac{\sum_{i=1}^{n}\left(y_{i}-\widehat{y_{i}} \right) ^{2}}{n}
\end{equation} 
In \cref{R2} and \cref{MSE}, $y_{i}$ is the real data, $\widehat{y_{i}}$ represents the predicted data, and $\overline{y}$ is the mean of the real data. In addition, $n$ is the number of samples, which in this study is the number of nodes (grids) in the computational domain.

\subsection{Hyperparameter Analysis}

\begin{figure}
    \centering
    \includegraphics[width=\columnwidth]{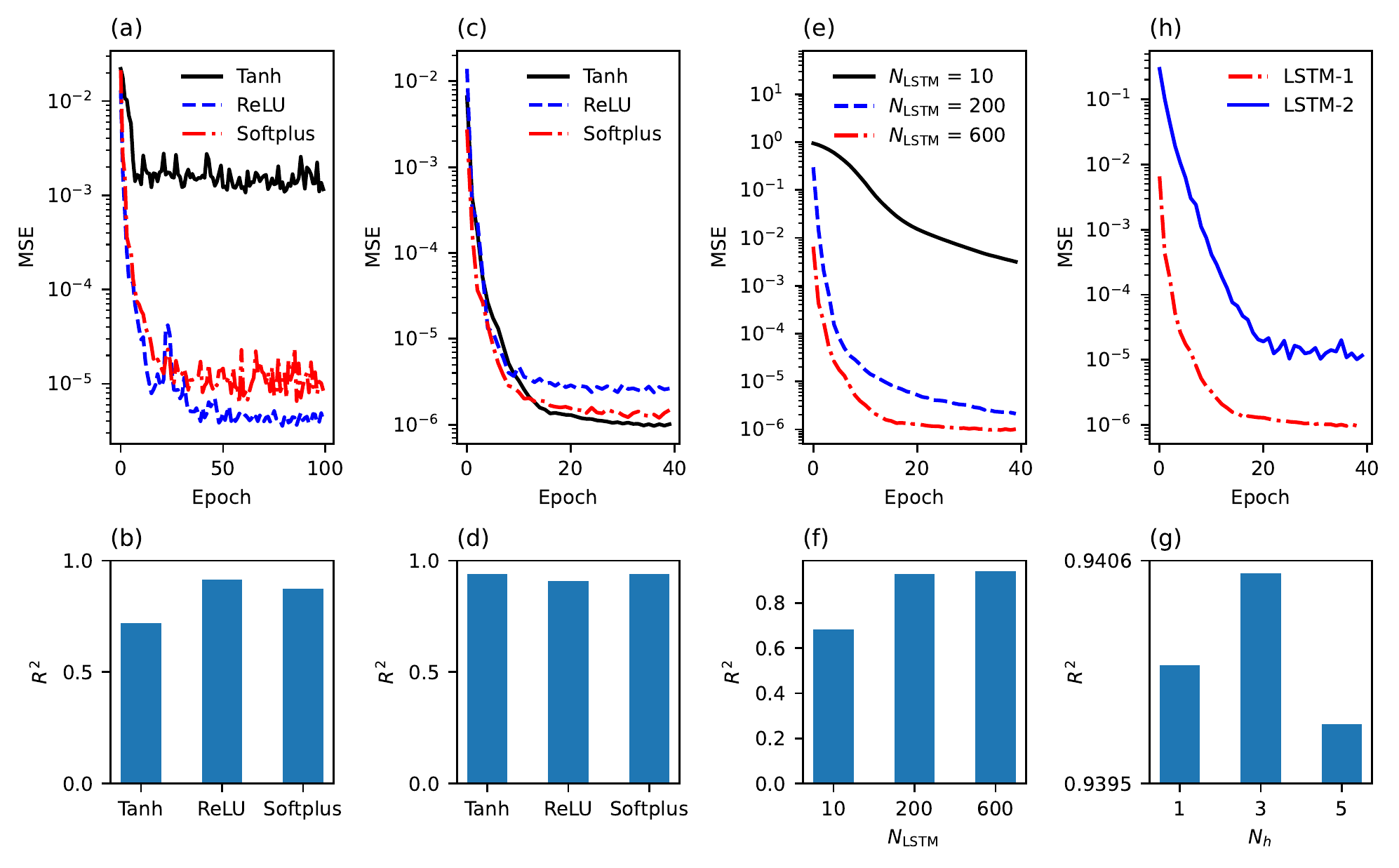}
        \caption{Results of the hyperparameter analysis; validation loss versus epochs (a) and $R^{2}$ of the reconstruction of the test data set (b) for autoencoder network with three different activation functions, validation loss versus epochs (c) and $R^{2}$ of the prediction of the test data set (d) for LSTM network with three different activation functions, validation loss versus epochs (e) and $R^{2}$ of the prediction of the test data set (f) for LSTM network with three different number of LSTM cells, effect of the number of hidden layers of the autoencoder network on the predictions with autoencoder-LSTM (g), and validation loss versus epochs for two different LSTM architectures (h)}.
        \label{hyperparams}
\end{figure}

The number of layers and the number of neurons in the hidden layers of both autoencoder and LSTM networks besides their activation functions are the hyperparameters of the presented method. An analysis of the hyperparameters is conducted, and the results for the most important ones are presented in \cref{hyperparams} for the cylinder test 2. Here, the performance of the autoencoder and LSTM networks in reconstruction and prediction of the testing data is investigated using three different activation functions, i.e., tanh, ReLU, and softplus, three different number of hidden layers for the autoencoder network (1, 2, 3), and three different number of LSTM cells (10, 100, 600) in the LSTM network. 15\% of the train data set is considered as the validation set. Results are presented as the validation loss during the training process and $R^{2}$ value of the prediction or reconstruction of the testing data. Note that error for the autoencoder network is the error in reconstruction of the flow-field from its dominant features, and the error for the LSTM is the prediction error. 

\Cref{hyperparams}\red{a} shows the validation loss versus number of epochs for the autoencoder network and \cref{hyperparams}\red{b} depicts the obtained $R^{2}$ for the test data set. It can be seen that the ReLU activation function performs better for the autoencoder. Here, results are reported for the autoencoder network with three hidden layers and $n_{g} = 50$. The first and third hidden layers contain 500 neurons. Note that we use linear activation functions for the Input, output, and the bottleneck layers and nonlinear activation functions for other layers. \Cref{hyperparams}\red{c} and \cref{hyperparams}\red{d} present the same results for the LSTM network indicating the better performance of the tanh activation function for the LSTM. Moreover, \cref{hyperparams}\red{e} and \cref{hyperparams}\red{f} shows the results for the LSTM networks with various number of LSTM cells. It can be seen that the increase in the number of LSTM cells leads to better predictions. Also, 
the effect of the number of hidden layers of the autoencoder in the accuracy of the autoencoder-LSTM predictions is investigated, and the results are reported in \cref{hyperparams}\red{g}. It can be observed that the autoencoder with three hidden layers obtains the best results. It also should be noted that to avoid overfitting, the training process is stopped where more training epochs do not lead to further reduction of the validation loss. Another choice for the architecture of the LSTM network is to consider $[g(t_{n}), ... , g(t_{n+p})]$ as the input and get $g(t_{n+p+1})$ at the output. Then, the velocity field $\tilde{V}(t_{n+p+1})$ can be obtained using the decoder part of the autoencoder network. We refer to this architecture as LSTM-2. We tested the performance of this model, and compare it with the architecture in which the LSTM network directly predicts the velocity field as described in $\S$\ref{method_autoLSTM}. We refer to this model as LSTM-1. We observed that lower validation loss for the velocity field can be obtained using the LSTM-1 architecture as it is shown in \cref{hyperparams}\red{h}. Note that for LSTM-2, we used the decoder part during the training of the LSTM to calculate the validation loss for the velocity field. Based on these results, the autoencoder network with three hidden layers and ReLU activation function and the LSTM-1 network with one hidden layer containing 600 LSTM cells with tanh activation function are chosen as the network architectures for the autoencoder-LSTM method. We also observed that these hyperparameters are robust for all the test cases, showing possible applicability of ReLU and tanh activation functions, respectively, for dimension reduction and future prediction in the domain of fluid dynamics.

\subsection{Structure of the Inputs and Outputs}

In this section, more detailed information on the shape of the inputs and outputs for each test case is presented. For the cylinder test cases, the snapshots are constructed from the velocity domain ($-2D<x<8D$ and  $-4D<y<4D$). This leads to a snapshot of 81401 nodes. In this regard, the input and output layers of the autoencoder have 81401 nodes. In the oscillating airfoil test case, snapshots are extracted from the velocity domain ($-2c<x<7c$ and  $-4c<y<4c$). Each snapshot consists of 99473 nodes, so the input and the output layers of the autoencoder network for this test case have 99473 nodes. Other properties of the network remain unchanged. The input data is normalized to cover both negative and positive values through the equation below:
\begin{equation}
\hat{V}_{i}= \frac{V_{i}- \mu_{V_{i}}}{\sigma_{V_{i}}},
\label{eq1}
\end{equation}
where $i$ represents each measured point, $V$ is the measured value, and $\mu_{V_{i}}$ is the time average and standard deviation of $V_{i}$. $\hat{V}$ represents the normalized value.

The inputs of the LSTM network are sequences of the outputs of the bottleneck layer of the autoencoder network $[g(t_{n}), ... , g(t_{n+p})]$ with the size of $n_{g}$. At the output, it is of interest to have the predicted velocity field. Therefore the output layer has the same number of nodes as the velocity field dimension, which is 81401 for the cylinder test cases and 99473 for the oscillating airfoil test case. The input of the LSTM network is a three-dimensional matrix indicating the size of the input arrays, $n_{g}$, the number of time steps in a sequence, $p$, and the number of sequences.

\subsection{DMD with Autoencoder for Dimensionality Reduction}

Dynamic mode decomposition is one of the well-known model-free reduced order modeling techniques that is based on the measurements of the system rather than the governing equations. In fluid dynamics, it is also a method to decompose complex flows into dominant spatiotemporal coherent structures using the power of the SVD. 
The first step for utilizing the DMD analysis is to construct the snapshots’ matrix $X$ and the lagged matrix $X'$:
\begin{equation}
    X=
    \left[ \begin{array}{ccccc}
        \mid & \mid & \mid & \mid \\ 
        x_{0} & x_{1} & \cdots & x_{m} \\  
        \mid & \mid & \mid & \mid 
    \end{array}\right],
\end{equation}

\begin{equation}
    X'=
    \left[ \begin{array}{ccccc}
        \mid & \mid & \mid & \mid \\ 
        x_{1} & x_{2} & \cdots & x_{m+1} \\  
        \mid & \mid & \mid & \mid 
    \end{array}\right].
\end{equation}
The locally linear approximation can be written in terms of these data matrices as below:
\begin{equation}
X' \approx A X.
\end{equation}
Therefore, the best-fit operator $A$ is given by:
\begin{equation}
A=X' X^{\dagger},
\end{equation}
where $\dagger$ is the Moore-Penrose pseudoinverse. The high-dimensional matrix $A$ is not computed directly; instead, it is first projected onto a low-rank subspace using SVD. The SVD of $X$ can be calculated as:
\begin{equation}
X = U \Sigma V^{*}.
\end{equation}
Then, the matrix $A$ is obtained from:
\begin{equation}
A= X' V\Sigma^{-1} U^{*}.
\end{equation}
Therefore, $\widetilde{A}$ the $r \times r$ projection of the matrix $A$ onto the low-ranked subspace can be obtained as below, where $r$ is the rank of truncation:
\begin{equation}
\widetilde{A} = U^{*} AU= U^{*} X' V \Sigma ^{-1}.
\end{equation}
Eigendecomposition of $\widetilde{A}$ obtain a set of eigenvectors $w$ and eigenvalues $\lambda$, where:
\begin{equation}
\widetilde{A}w = \lambda w.
\end{equation}
For each pair of $w$ and $\lambda$, a DMD eigenvalue, which is $\lambda$ itself, and a DMD mode can be defined as:
\begin{equation}
\phi = \frac{1}{\lambda} X' V \Sigma^{-1} w.
\end{equation}
Finally, the approximate solution of at the future times can be calculated from:
\begin{equation}
x(t)  \approx \sum_{k=1}^{r} \phi_{k} \ e^{(\omega_{k} \ t)} \ b_{k},
\end{equation}
where, $\omega_{k} = \ln (\frac{\lambda_{k}}{\Delta t})$ and $b = \Phi^{\dagger} x_{1}$ is the initial amplitude of each mode. $x_{1}$ is the initial snapshot, and $\Phi$ is the matrix of DMD eigenvectors. 

\begin{figure*}[hbtp]
    \centering
    \includegraphics[width=0.8\textwidth]{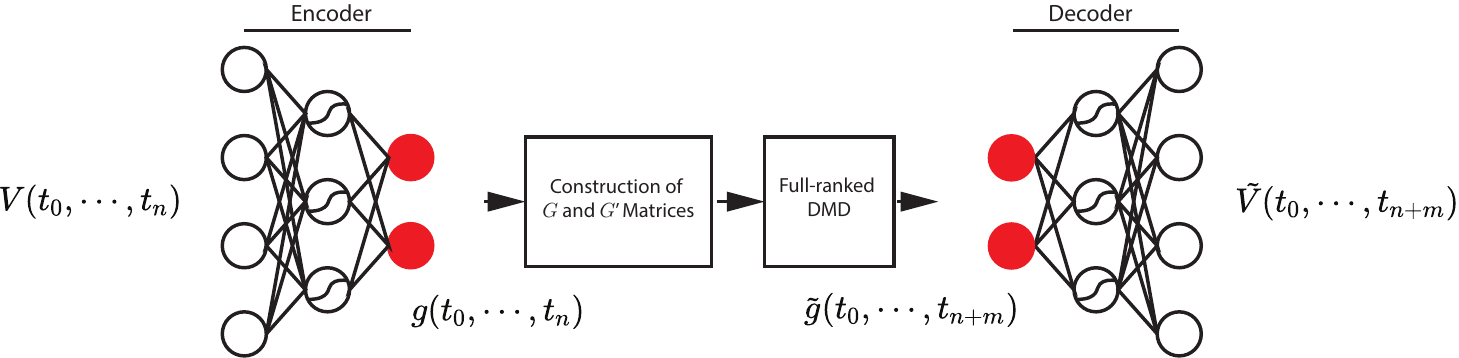}
    \caption{Schematic of DMD with autoencoder for dimensionality reduction; here, $m$ is the number of prediction steps}
    \label{DMDauto}
\end{figure*}

\begin{figure}[hbtp]
    \centering
    \includegraphics[width=0.8\columnwidth]{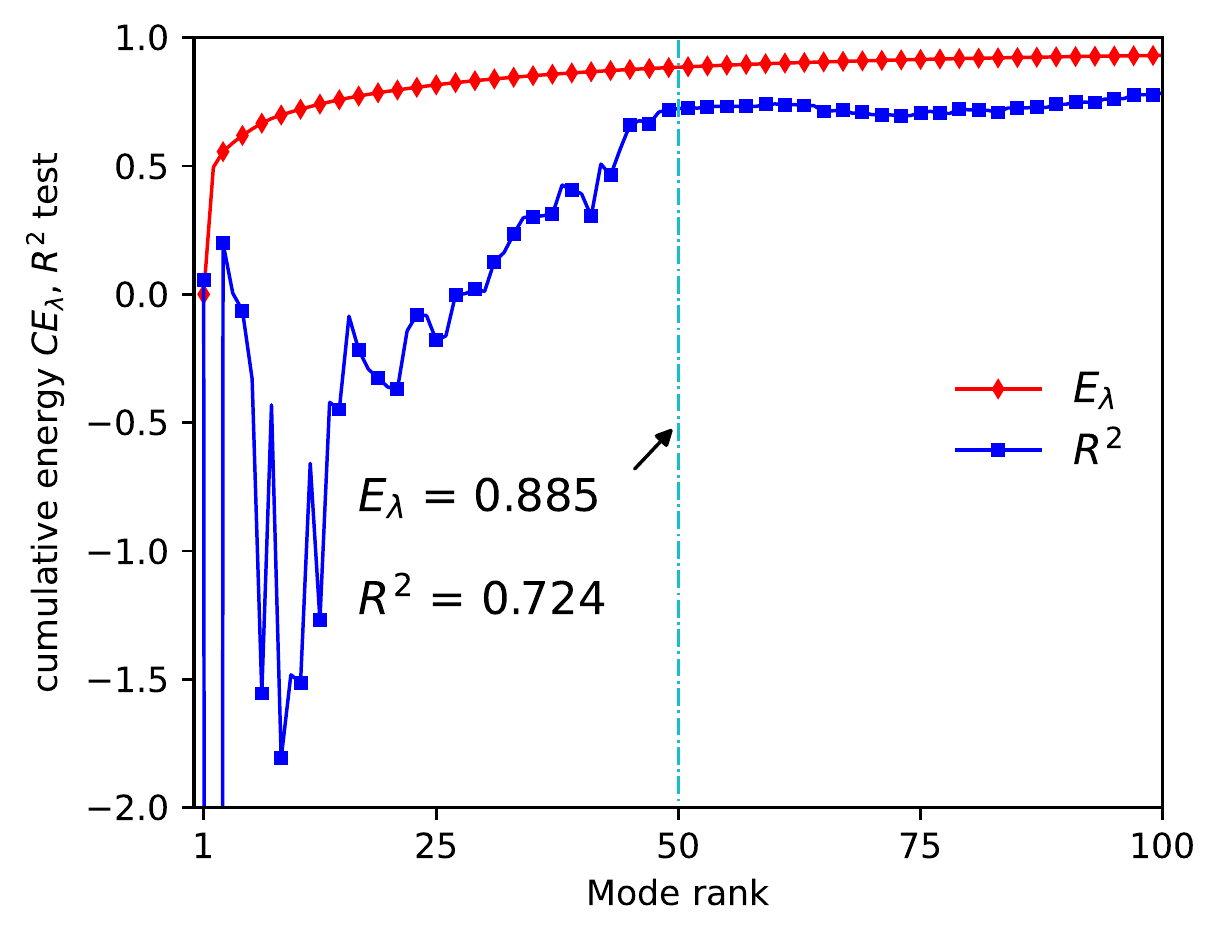}
    \caption{Cumulative energy of the DMD modes and coefficient of determination ($R^{2}$) for the oscillating airfoil}
    \label{DMD_modes_cumulative_energy}
\end{figure}

In this work, we also investigate the possibility of performing DMD analysis with the autoencoder network as a tool for dimension reduction rather than SVD. The velocity field around the test cases is reduced to the latent space of $g$. These reduced order vectors can be interpreted as the input snapshots, allowing to construct the input matrices required for the DMD as:
\begin{equation}
    \begin{split}
        &G = [g(t_{0})~~ g(t_{1})~~ \cdots ~~g(t_{m})],\\
        &G^{\prime} = [g(t_{1})~~ g(t_{2})~ \cdots ~~ g(t_{m+1})].
    \end{split}
\end{equation}
Then, a full-ranked DMD analysis can be performed on the matrices $G$ and $G^{\prime}$ to compute the mapping matrix of $A = G^{\prime}G^{\dagger}$, which can propagate the data into the future time steps. After performing DMD on the $g$ space, the predicted values are passed from the decoder part, $\psi$, of the autoencoder network to obtain the predicted velocity field for future time steps. \Cref{DMDauto} illustrates the schematic of this method, which is referred, hereafter, as autoencoder-DMD. In this study, DMD analysis is performed with the use of the PyDMD \cite{demo18pydmd} library in Python. It is worth to mention that for choosing the number of DMD modes, we follow a simple procedure. As shown in \cref{DMD_modes_cumulative_energy} for the oscillating airfoil, the cumulative energy of DMD modes and the accuracy of the test case ($R^{2}$) are flattened at a certain point and converge to a specific number. In this regard, we have used this criteria for choosing the number of DMD modes for prediction. In other words, the number of DMD modes is selected when the accuracy of our method reaches a particular threshold. In this way, we ensure that the DMD method has the highest efficiency with the optimum number of parameters that are not affected by any sub-optimal or erroneous modes.

\section{Results and Discussion}

\subsection{Dimension Reduction with Autoencoder}

The autoencoder network has been used as a nonlinear dimension reduction technique to project the high dimensional data onto the low dimensional subspace. The velocity field is processed through the lower-dimensional representation at the bottleneck layer and then mapped back to the ambient dimension. Data loss is an inevitable consequence of dimension reduction, and the amount of compression which leads to a reasonable approximation error in the reconstructed data is significant. Therefore, the number of neurons at the bottleneck layer of the autoencoder network, $n_{g}$, is an important hyperparameter and should be chosen based on thorough experiments. Here, autoencoder networks with the size of the bottleneck layer of 3, 5, 25, and 50 are trained and tested with the training and testing data sets of the three test cases. $R^{2}$ of the testing data is reported in Table \ref{TableDR}. It can be seen that even with $n_{g} = 3$, the data has been reconstructed with acceptable accuracy. As it is expected, an increase of the dimension at the bottleneck layer leads to an increase in the accuracy of the reconstruction and a decrease of the approximation error.

\begin{center}
    \begin{table}[hbtp]
    \caption{Approximation errors of the dimension reduction using autoencoder networks with various number of neurons at the bottleneck layer}\label{TableDR}
    \centering
    \begin{tabular}{cccc}
        \hline 
        $n_{g}$  &  Cylinder test 1 &  Cylinder test 2 & Pitching airfoil  \\
     \hline
        \rule{0pt}{3ex}  3 & 0.999  & 0.882 & 0.927 \\
      5 & 0.999  & 0.899  & 0.930  \\
      25 & 0.999  & 0.928  & 0.932 \\
      50 & 0.999  & 0.936 & 0.936 \\
        \hline
    \end{tabular}
    \end{table}
\end{center}

At the latent space, the autoencoder network extracts the main features of the flow-field, which is adequate for the reconstruction of the velocity field at the output layer. To visualize the main features of the flow for the aforementioned test cases, an input array of $\mathfrak{g}$ is given to the bottleneck layer, and the results are taken from the output layer. For visualizing the features extracted by the first neuron, $\mathfrak{g}(1)$ is considered equal to one and others, $\mathfrak{g}(2), \cdots, \mathfrak{g}(n_g)$, are equal to zero. For the second neuron, $\mathfrak{g}(2)$ is equal to one and others are zero, and so on. 

\begin{figure}[thbp]
    \centering
    \includegraphics[width=\columnwidth]{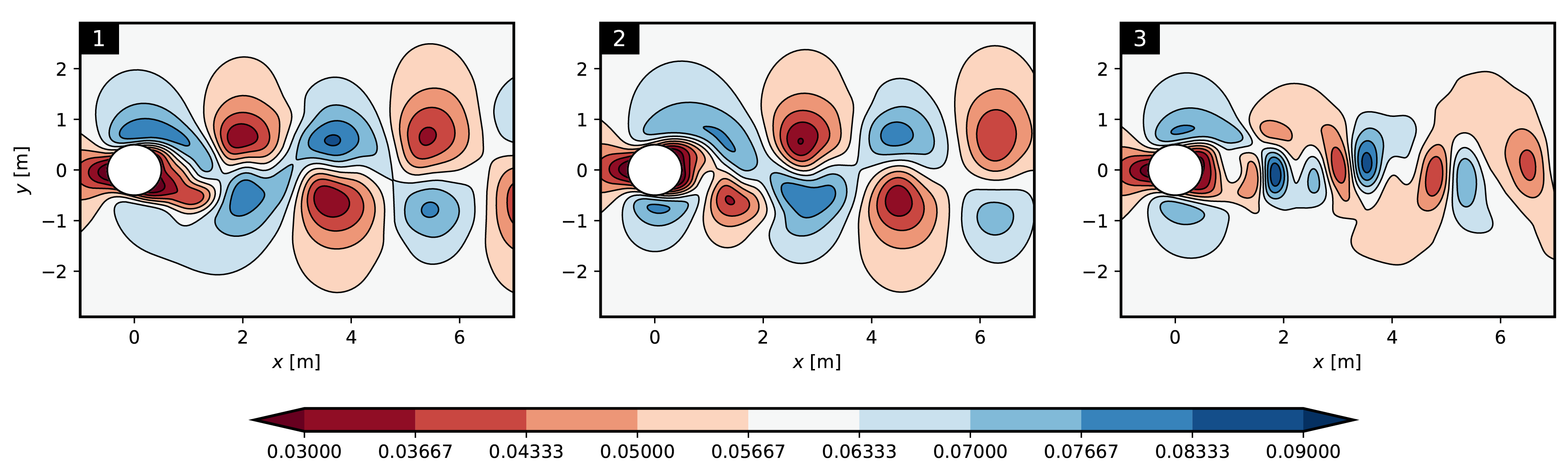}
    \caption{Autoencoder modes; cylinder test 1}
    \label{automodestst1}
\end{figure}

\begin{figure}[hbtp]
    \centering
    \includegraphics[width=\columnwidth]{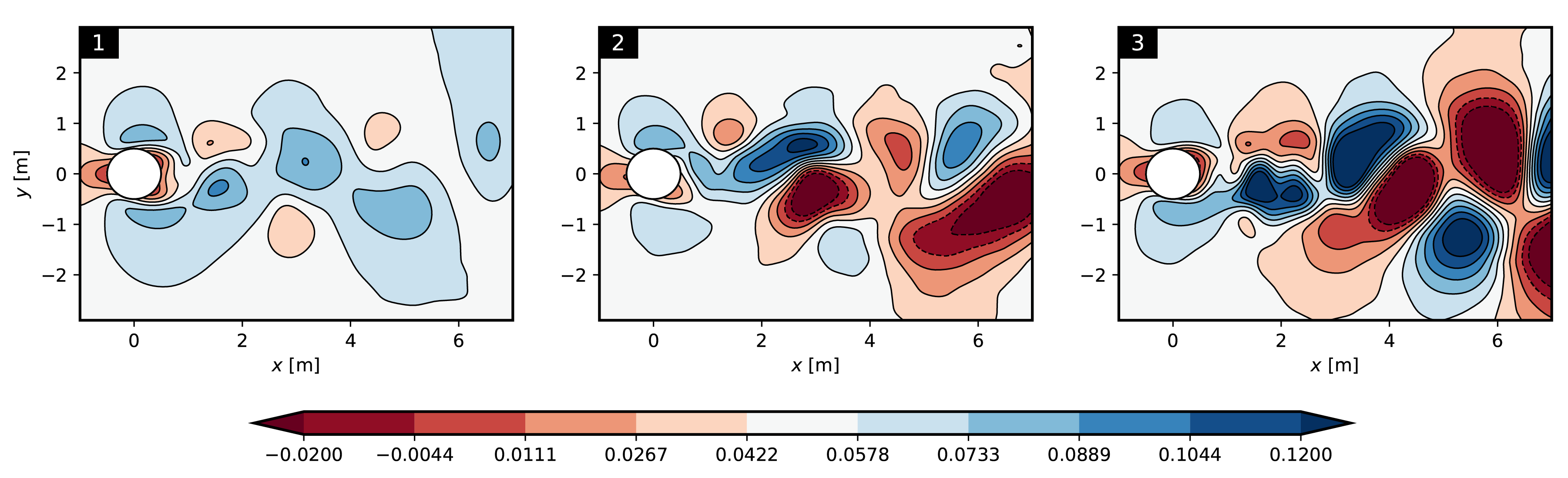}
    \caption{Autoencoder modes; cylinder test 2}
    \label{automodestst2}
\end{figure}

\begin{figure}[hbtp]
    \centering
    \includegraphics[width=\columnwidth]{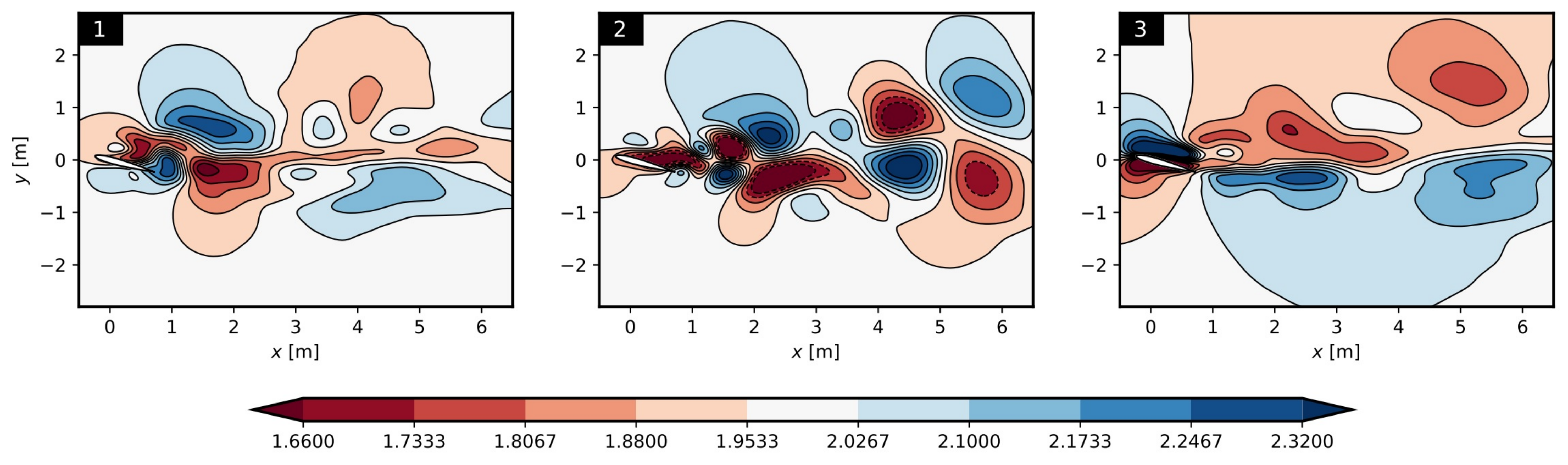}
    \caption{Autoencoder modes; pitching airfoil}
    \label{automodesairfoil}
\end{figure}

\begin{figure}[hbtp]
    \centering
    \includegraphics[width=\columnwidth]{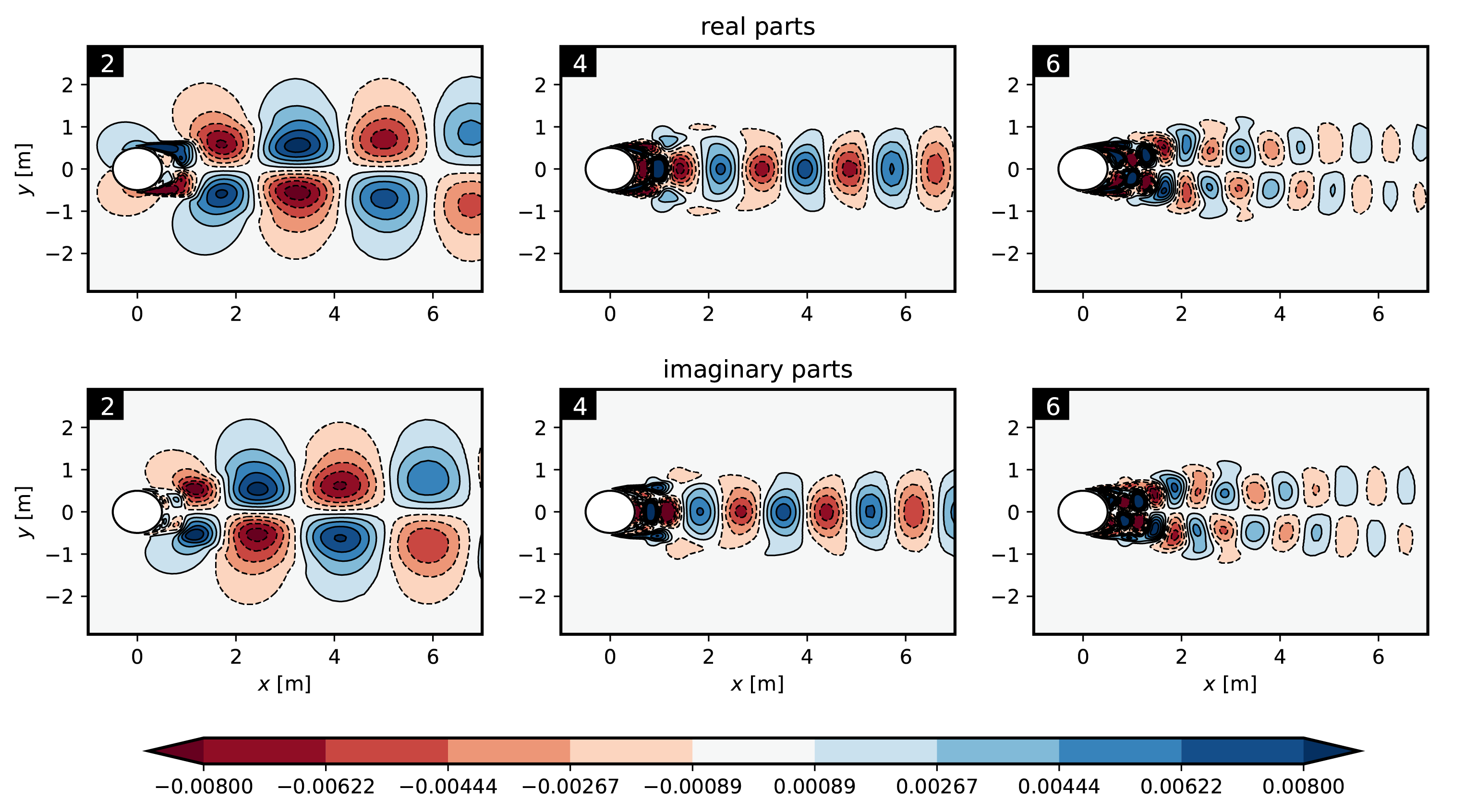}
    \caption{DMD modes; cylinder test 1}
    \label{DMDmodestst1}
\end{figure}

\begin{figure}[hbtp]
    \centering
    \includegraphics[width=\columnwidth]{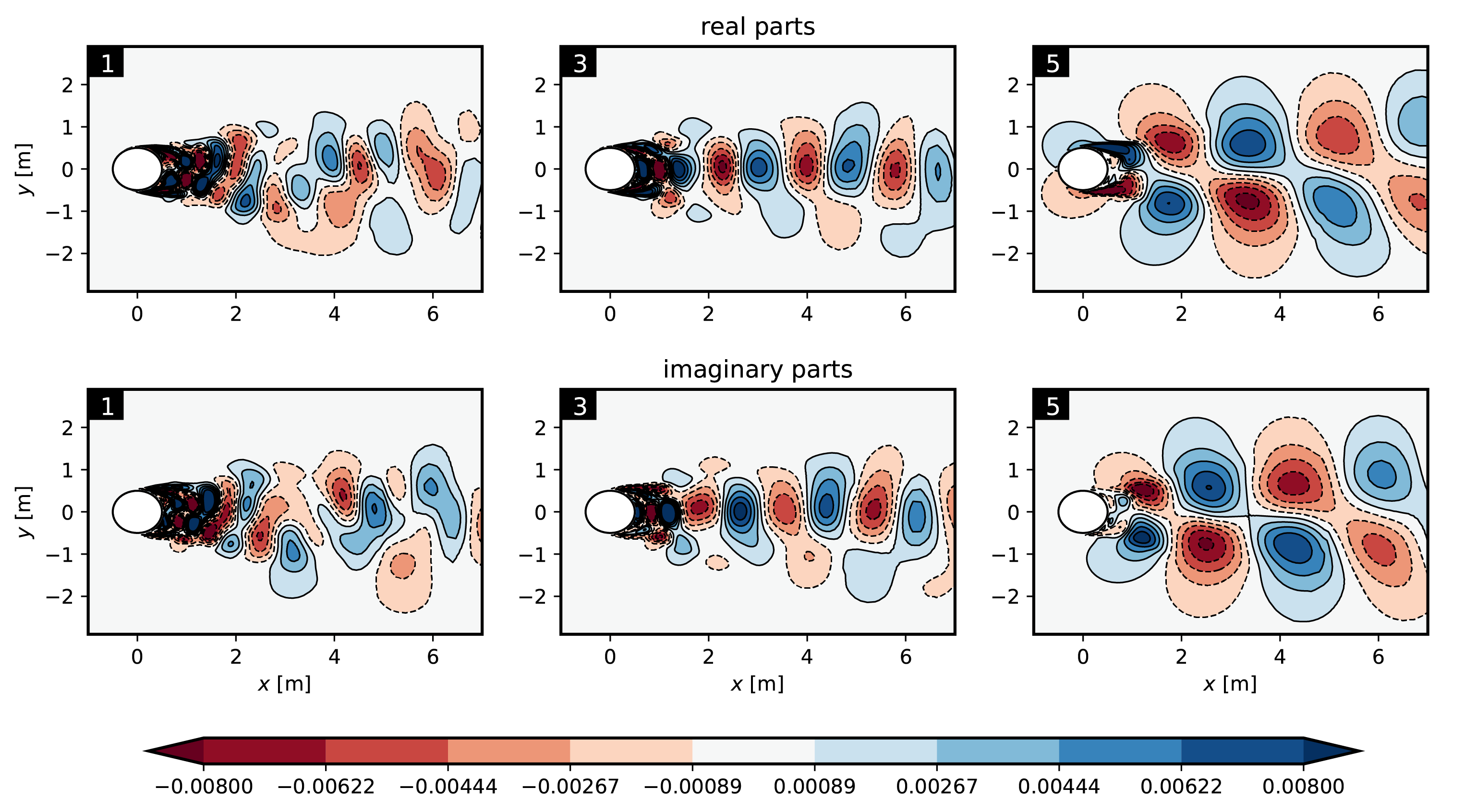}
    \caption{DMD modes; cylinder test 2}
    \label{DMDmodestst2}
\end{figure}

\begin{figure}[hbtp]
    \centering
    \includegraphics[width=\columnwidth]{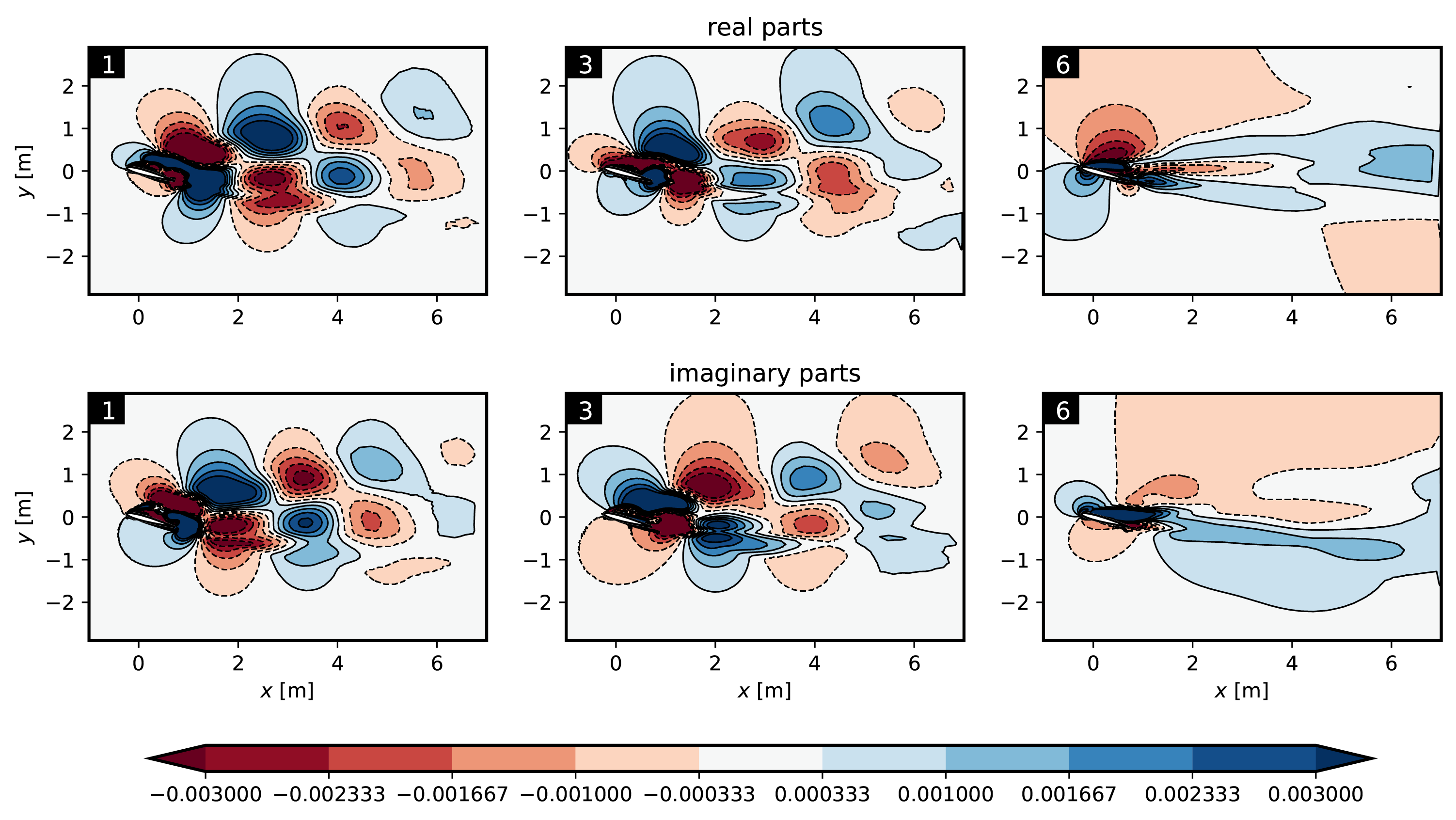}
    \caption{DMD modes; pitching airfoil}
    \label{DMDmodesairfoil}
\end{figure}

\Cref{automodestst1,automodestst2,automodesairfoil} illustrate the first three features extracted by the autoencoder network  (autoencoder modes) from the velocity field for the cylinder test 1, cylinder test 2, and the pitching airfoil, respectively. \Cref{DMDmodestst1,DMDmodestst2,DMDmodesairfoil} depict the DMD modes for the mentioned test cases.
It can be seen that the autoencoder modes are different from the modes of decomposition-based model reductions, such as DMD modes. The main reason is the nonlinearity added to the dimension reduction technique using nonlinear activation functions. However, similar to the DMD modes, the autoencoder modes represent the most dominant features of the flow-field. For better representation, DMD modes corresponding to the steady mean flow and the complex conjugate modes are not illustrated. We observed that the extracted features from the autoencoder may contain noise, and we performed a two-dimensional Gaussian filter to provide a better representation of the dominant modes.

\subsection{Comparison of Autoencoder-LSTM with DMD-based models}
\label{auto-lstm-sec}

Here, the performance of the proposed ROM framework, autoencoder-LSTM, in the prediction of the future estate of the test cases is compared with the DMD and autoencoder-DMD methods. Results are reported for the train and test data sets, where the train data set is reconstructed and the test data set is predicted by the models. We report the performance of the models for the training data sets to also show their skills in the reconstruction of the flow-field from its dominant modes. However, the main focus is on the comparison of the prediction skills of the models. For all the models, the size of the latent space is equal to 50, which means that $n_{g} = 50$ for the autoencoder network, and $r = 50$ for the DMD.

\begin{center}
    \begin{table}[hbtp]
    \caption{$R^{2}$ and MSE of the examined models for the prediction and reconstruction of the testing and training data sets. The latent space dimension $n_g$ is equal to 50.}\label{tmse}
    \centering
    \resizebox{\columnwidth}{!}{%
    \begin{tabular}{@{\extracolsep{5pt}}cccccc@{}}
    \hline
    \multirow{ 2}{*}{Test cases} & & \multicolumn{2}{c}{Autoencoder-LSTM} & \multicolumn{2}{c}{DMD} \\ 
    \cline{3-4} \cline{5-6}                          
    \rule{0pt}{3ex} & & $R^{2}$     & MSE    &  $R^{2}$     & MSE \\ 
      \hline
      \rule{0pt}{3ex} \multirow{ 2}{*}{Cylinder 1}&Train &0.9999&$2.70\times 10^{-9}$&0.9982& $1.02\times 10^{-6}$ \\ 
      &Test                &0.9988&$8.46\times 10^{-9}$&0.9949& $2.93\times 10^{-6}$ \\
      \rule{0pt}{4ex} \multirow{ 2}{*}{Cylinder 2}&Train &0.9999&$1.13\times 10^{-8}$&0.9937& $4.58\times 10^{-7}$ \\ 
      &Test                &0.9405&$1.06\times 10^{-6}$&0.0826& $2.00\times10^{-5}$ \\
      \rule{0pt}{4ex} \multirow{ 2}{*}{Oscillating airfoil}&Train &0.9431&$3.58\times 10^{-4}$&0.7796& $3.59\times 10^{-2}$ \\ 
      &Test                &0.9421&$6.17\times 10^{-4}$&0.7239& $4.64\times 10^{-2}$ \\
      \hline
    \end{tabular}
    }
    \end{table}
\end{center}

$R^{2}$ and MSE obtained from the models in the prediction of the training and testing data sets are reported in \cref{tmse}. It can be seen that the autoencoder-LSTM method obtains the $R^{2}$ of at least 0.9405 in the prediction of the test data set of the cases, which indicates the excellent performance of this method in prediction of the future state of the flow, only from past measurements. For the cylinder test 1, the DMD performance is comparable with the autoencoder-LSTM. However, for the cylinder test 2 and oscillating airfoil, which exhibit, respectively, multi-frequency and extreme events phenomena, the performance of the autoencoder-LSTM method outperforms the DMD method. 

\begin{figure}[hbtp]
    \centering
    \includegraphics[width=\columnwidth]{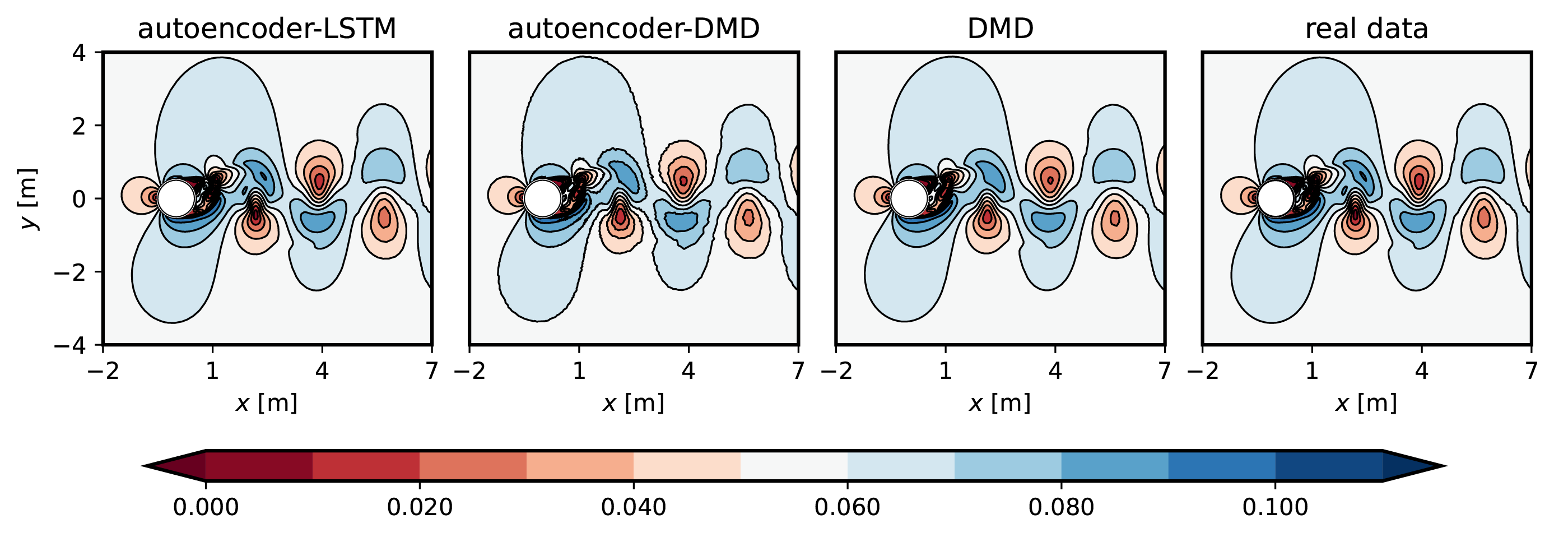}
    \caption{Contour of velocity magnitude over the cylinder test 1; prediction of the various models against the real data at the last prediction step}
    \label{test1_c}
\end{figure}

\begin{figure*}[hbtp]
    \centering
    \includegraphics[width=0.7\textwidth]{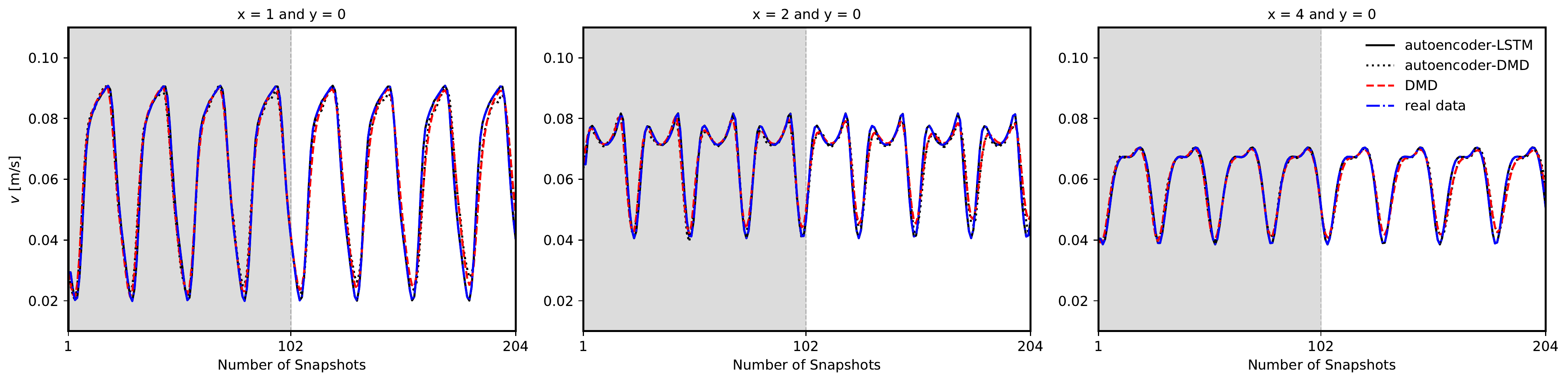}
    \caption{Time evolution of the velocity magnitude at three different points in the wake of the cylinder test 1; predictions of the various models against the real data}
    \label{test1_l}
\end{figure*}

\Cref{test1_c} shows the velocity magnitude over the cylinder test 1 at the last prediction step for the various models against the real data. Moreover, to depict the performance of the models in the prediction of the velocity evolution through the time, the data for three different points in the wake of the cylinder at $x = 1, 2, 4$ and $y=0$ are presented in \cref{test1_l}. The grey area shows the training data that is reconstructed by the models, and the white area shows the predictions of the testing data. It can be seen that for this test case, which is a periodic dynamical system, all of the data-driven models provide accurate results. The best predictions of the test data set are obtained from the autoencoder-LSTM method leading to the $R^2$ of 0.9988 against 0.9949 of the DMD.

\begin{figure*}[hbtp]
    \centering
    \includegraphics[width=0.7\textwidth]{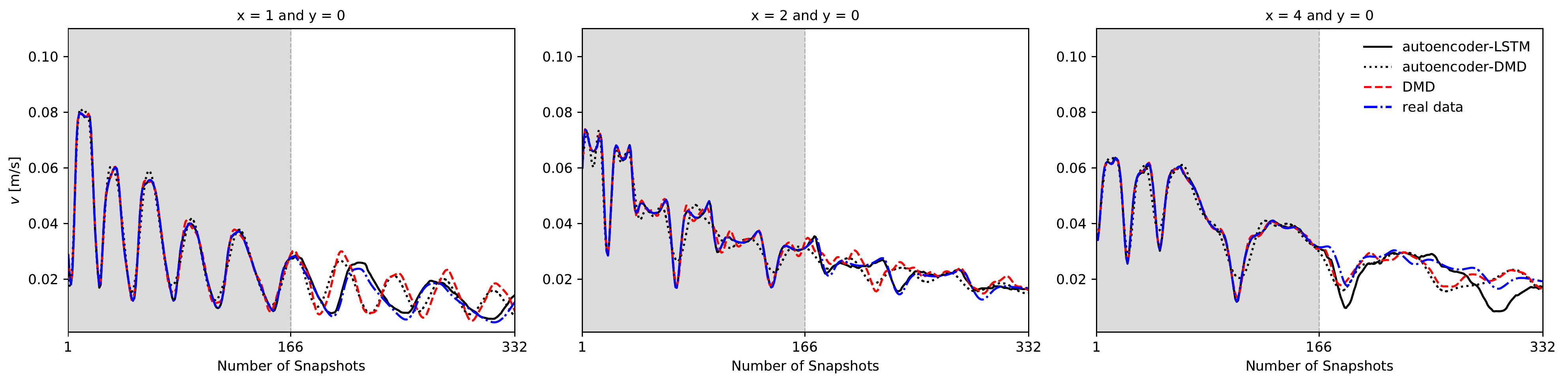}
    \caption{Time evolution of the velocity magnitude at three different points in the wake of the cylinder test 2; predictions of the various models against the real data}
    \label{test2_l}
\end{figure*}

For the cylinder test 2, the autoencoder-LSTM method performs much better than the DMD method in prediction of the velocity variations through the time acquiring $R^{2}$ of 0.9405 against 0.0826 of the DMD for the testing data (see \Cref{tmse}). \Cref{test2_l} represents the time evolution of the velocity magnitude at three different points in the wake of the cylinder test 2. Again, the grey area shows the data which is inside the training data set, and the white area shows the data which is outside of the training data set. Here, It can be seen that the DMD method is not able to predict the time variations of the velocity for the testing data and leads in variations with higher frequencies while the autoencoder-LSTM method performs very well in the prediction of the flow evolution. It also can be observed that the performance of the autoencoder-DMD is almost the same as the DMD method.

\begin{figure}[hbtp]
    \centering
    \includegraphics[width=\columnwidth]{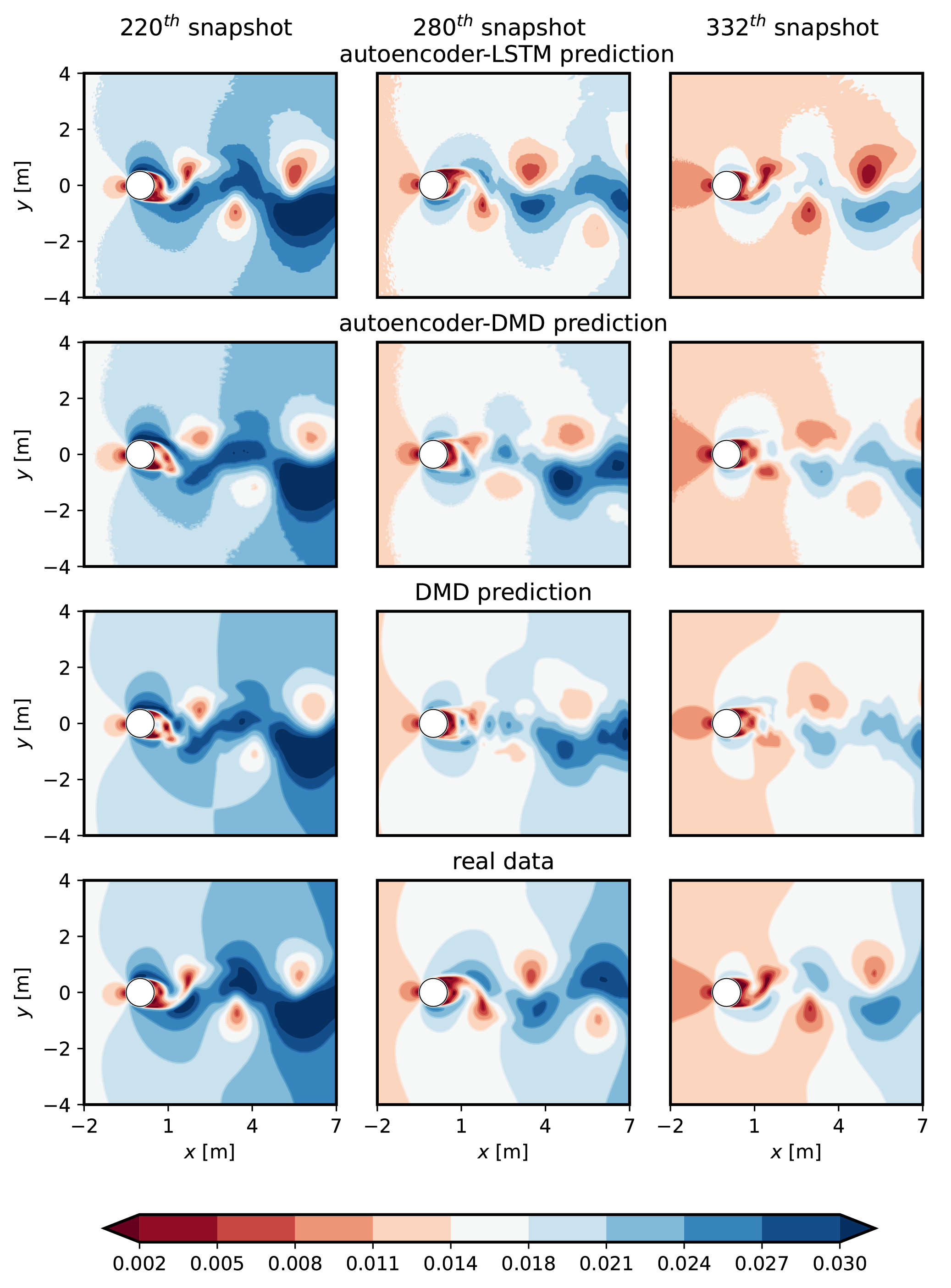}
    \caption{Contours of velocity magnitude over the cylinder test 2; prediction of the various models against the real data for three time instances corresponding to 220$^{th}$, 280$^{th}$ and 332$^{th}$ snapshots}
    \label{test2_c}
\end{figure}

To illustrate the performance of the models in the prediction of the flow-field, contours of velocity magnitude are shown in \cref{test2_c} at three time instances corresponding to the three snapshots at the testing data. It can be seen that the autoencoder-LSTM method, can acquire trustworthy results in the prediction of the time evolution of a complex fluid system while the well-known method of DMD is not able to predict the future time instances properly for this test case since it is essentially a linear model. We compared the performance of the proposed autoencoder-LSTM model with the DMD in the prediction of this non-periodic test case to give an impression of the importance of modeling of nonlinear processes. The inaccuracy of the DMD method is not just related to the SVD rank truncation, where a full ranked DMD analysis of the cylinder test 2 leads to $R^{2}$ of 0.1776 and MSE of $1.81\times10^{-5}$ for the testing data; however, autoencoder-LSTM provides $R^{2}$ of 0.9405 and MSE of $1.06\times10^{-6}$. The accuracy of the DMD results may be enhanced by the increase of the number of snapshots taken from the flow-field, but here, it can be concluded that the autoencoder-LSTM method provides more accurate results than the DMD method for an identical data set.

\begin{figure*}[hbtp]
    \centering
    \includegraphics[width=0.7\textwidth]{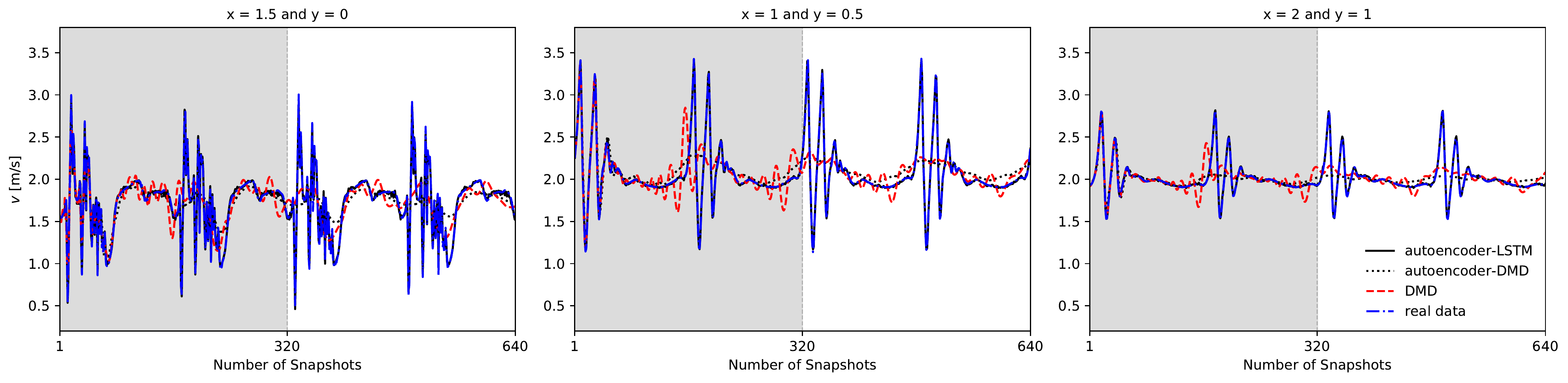}
    \caption{Time evolution of the velocity magnitude at three different points in the wake of the oscillating airfoil; predictions of the various models against the real data}
    \label{test3_l}
\end{figure*}

In the oscillating airfoil test case, dynamic stall occurs when the airfoil's angle of attack surpasses a specific angle. The dynamic stall is a complex nonlinear phenomenon that follows with strong variations of the flow parameters around the airfoil due to the consecutive vortex shedding. By the increase of the angle of attack, the boundary layer separates from the airfoil surface leading to the dynamic stall phenomena, and by the decrease of the angle of attack through the airfoil oscillations, the boundary layer reattaches to the airfoil surface. Here, it is of value to assess the ability of the data-driven ROMs in the prediction of the dynamic stall. \Cref{test3_l} shows the variation of the velocity magnitude versus time for the oscillating airfoil at three different points in the wake. Strong variations in velocity magnitude indicate the dynamic stall and consecutive vortex shedding. It can bee seen that the autoencoder-LSTM predicts the velocity variations accurately, which shows the excellent performance of this method in forecasting of an extreme event. The DMD and the autoencoder-DMD methods, however, are not able to perfectly capture the dynamic stall phenomena, and it can be seen that even the reconstruction of the train data set is not accurate. 

\begin{figure}[hbtp]
    \centering
    \includegraphics[width=\columnwidth]{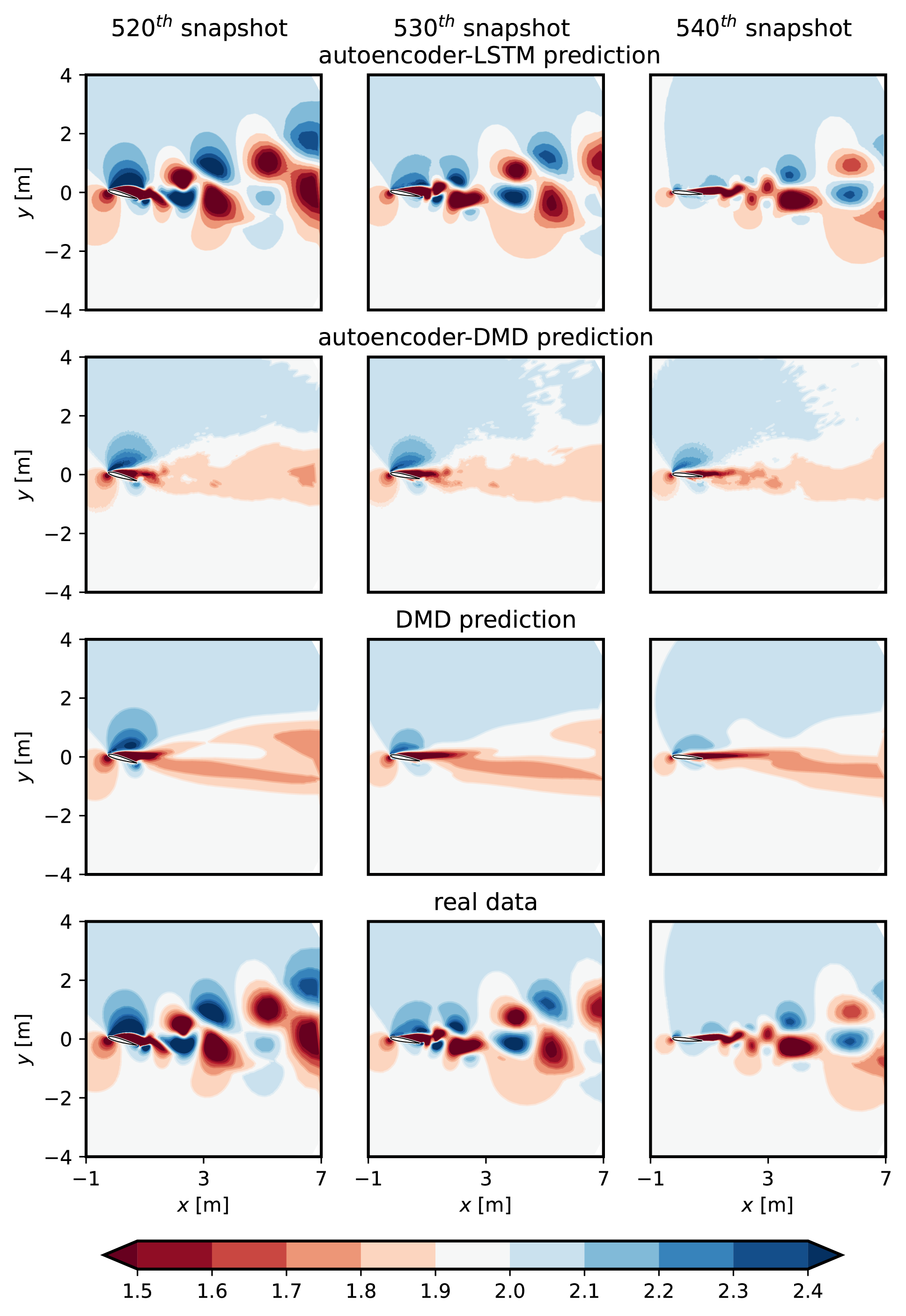}
    \caption{Contours of velocity magnitude over the oscillating airfoil; prediction of the various models against the real data for three time instances after occurrence of the dynamic stall}
    \label{test3_c}
\end{figure}

To provide a better insight into the physics of the dynamic stall and the performance of various models in prediction of the flow-field, \cref{test3_c} presents the contours of velocity magnitude over the airfoil at three time instances after occurrence of the dynamic stall. Here, the excellent performance of the autoencoder-LSTM against the DMD and autoencoder-DMD methods is evident. It can be seen that the vortex shedding behind the airfoil is very well predicted by the autoencoder-LSTM while the predictions of the other methods are inaccurate. Our results show the excellent performance of the proposed autoencoder-LSTM method in both dimension reduction and future prediction of the periodic and non-periodic flows investigated in this paper. However, the viability of this framework for very large-scale test cases with hundreds of millions nodes should be investigated through more general model assessment.

\subsection{Comparison of Autoencoder-LSTM with POD-based models}
In \cref{auto-lstm-sec}, we compared autoencoder-LSTM method with linear non-intrusive ROMs based on Koopman operator to show the effect of adding nonlinearities on the performance of the reduced order model. It is also interesting to compare results with POD-based ROMs. A conventional strategy is to use the spatial POD functions, the so-called POD modes, as basis functions for Galerkin method to construct a system of ordinary differential equation (ODEs), known as Galerkin system (GS), which approximates the temporal dynamics. Since fluid flows are often consisted only a small number of coherent structures, a low-order model can be constructed using a few dominant POD modes, leading to an ODE system of small dimension. However, the obtained ODE system may be inaccurate and unstable especially for transitional and turbulent flows \cite{Rempfer2000}. Moreover, it may be challenging to represent highly nonlinear problems, such as high Reynolds number flows, only with few POD modes due to the linear nature of the method. For instance, 7260 most energetic modes are needed to reconstruct about 95\% of the total energy content of a turbulent channel flow at $Re_{\tau} = 180$ \cite{Alfonsi}. Recently, efficient strategies have been proposed for non-intrusive ROM based on POD modes by utilizing DNNs and RNNs for temporal predictions of the amplitudes and to overcome the limitations of the Galerkin method \cite{Wang2018,Mohan2018,Pawar2019,pawarDatadrivenRecoveryHidden2020}. However, these methods may still suffer from fundamental challenges of traditional POD-based models. Here, we compare the autoencoder-LSTM method with a POD-based method which uses the LSTM network for temporal predictions of the mode amplitudes $a(t)$. We refer to this model as POD-LSTM. Note that for the POD-LSTM method, the LSTM network predicts the temporal dynamics in the low-order dimension while for the autoencoder-LSTM, it directly predicts the temporal evolution of the flow from the past low-order representations. We also will discuss the viability of performing predictions in the low-order dimension for the autoencoder-LSTM later in this section.
\begin{figure*}[hbtp]
    \centering
    \includegraphics[width=\textwidth]{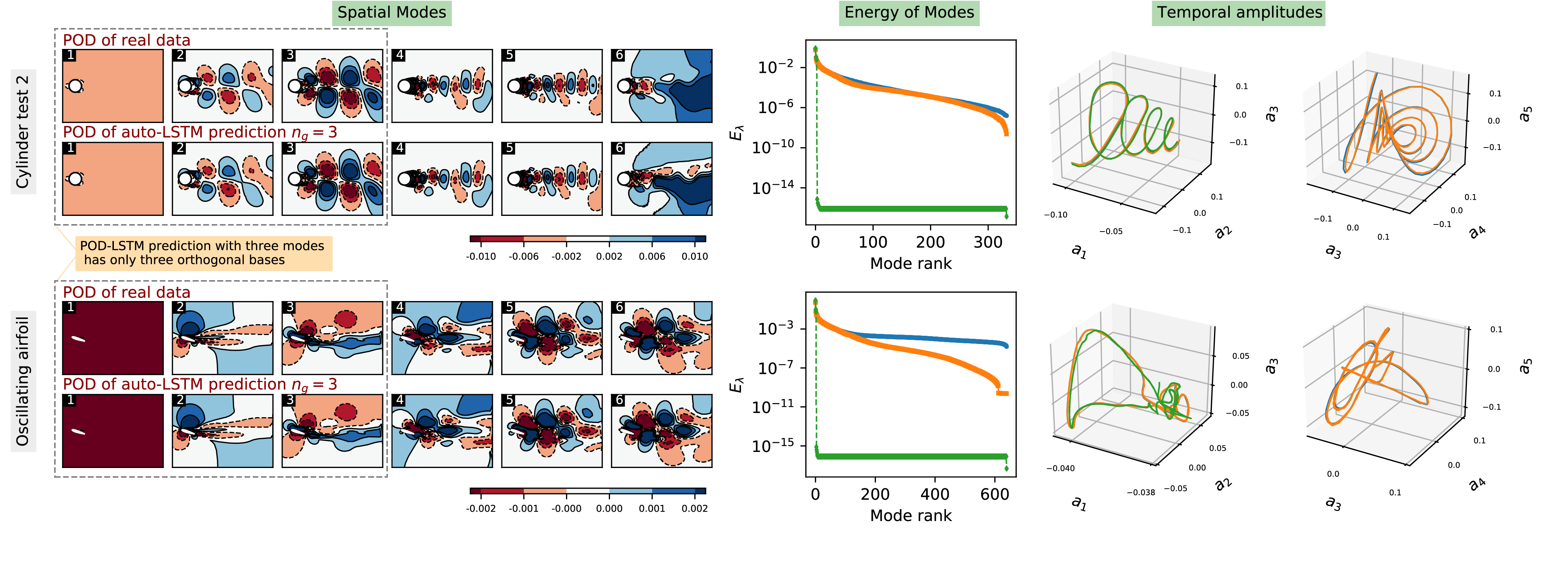}
    \caption{POD of the predicted velocity fields and the original reference data. Blue lines represent the results for the real data, and orange and green shows the results obtained from the predictions of the autoencoder-LSTM and POD-LSTM methods, respectively.}
    \label{pod}
\end{figure*}

We tested the performance of the models for two latent space dimension $n_{g}$ of 3 and 50. For both methods, the LSTM network contains one hidden layer with 600 LSTM cells and tanh activation function. For the cylinder test 1, both methods provide excellent predictions with $R^{2}$ of above 0.99, so here, we only reported the results for the cylinder test 2 and the airfoil as more challenging test cases. In the first step, it is interesting to evaluate the effect of the efficiency of dimension reduction technique on the accuracy of the predictions. To this end, we predict the whole time span of each test case using models with the latent dimension of 3 (POD rank truncation of 3 for the POD-LSTM and 3 neurons at the bottleneck layer for the autoencoder-LSTM) and obtain the predicted velocity snapshots. We then perform a POD of the predicted snapshots to see how many POD bases are captured in the predictions. \Cref{pod} presents the results of the decomposition for the predicted and the real data for both test cases of the cylinder test 2 and the oscillating airfoil. For the POD-LSTM method, the predictions based on 3 modes only contains 3 POD basis as is expected. These modes are the same as the ones for the real data, so they are not depicted in \cref{pod}. It can be seen that the energy of the modes drops to almost zero after mode 3. The interesting view is that the LSTM network perfectly exploits the dynamics of the temporal amplitudes of these 3 modes for both test cases. In this case, the temporal amplitudes for higher modes are not depicted for POD-LSTM since they are just noise.

For the autoencoder-LSTM, the interesting results is that the predictions based on only 3 autoencoder modes contain more than 3 POD basis as it can be observed in \cref{pod} for up to mode 6. In fact, the autoencoder can compact a higher number of POD modes in just 3 autoencoder modes. Moreover, the energy of the modes are comparable to the ones from the original data especially for the modes with higher energy. The LSTM network also can provide excellent predictions of the temporal amplitudes as can be seen in \cref{pod}. In \cref{tab:pod} we report $R^2$ and MSE of the predictions and reconstructions of the train and test sets. Our results show that excellent predictions can be obtained from the autoencoder-LSTM even using $n_g$ of 3 due to the promising ability of the autoencoder in dimension reduction.

\begin{center}
    \begin{table}[hbtp]
    \caption{$R^{2}$ and MSE of the autoencoder-LSTM and POD-LSTM models in the prediction and reconstruction of the original test and train data. The latent space dimension $n_g$ is equal to 3.}\label{tab:pod}
    \centering
    \resizebox{\columnwidth}{!}{%
    \begin{tabular}{@{\extracolsep{5pt}}cccccc@{}}
    \hline
    \multirow{ 2}{*}{Test cases} & & \multicolumn{2}{c}{Autoencoder-LSTM} & \multicolumn{2}{c}{POD-LSTM} \\ 
    \cline{3-4} \cline{5-6}                          
    \rule{0pt}{3ex} & & $R^{2}$     & MSE    &  $R^{2}$     & MSE \\ 
      \hline
      \multirow{ 2}{*}{Cylinder 2}&Train &0.9998&$1.34\times 10^{-8}$&0.7148&$4.55\times 10^{-6}$ \\ 
      &Test&0.9095& $1.58\times 10^{-6}$&0.6988& $4.72\times10^{-6}$ \\
      \rule{0pt}{4ex} \multirow{ 2}{*}{Oscillating airfoil}&Train &0.9434&$2.77\times 10^{-4}$&0.6416& $5.32\times 10^{-2}$ \\ 
      &Test&0.9424&$5.16\times 10^{-4}$&0.6403& $5.32\times 10^{-2}$ \\
      \hline
    \end{tabular}
    }
    \end{table}
\end{center}

\begin{figure}[hbtp]
    \centering
    \includegraphics[width=\columnwidth]{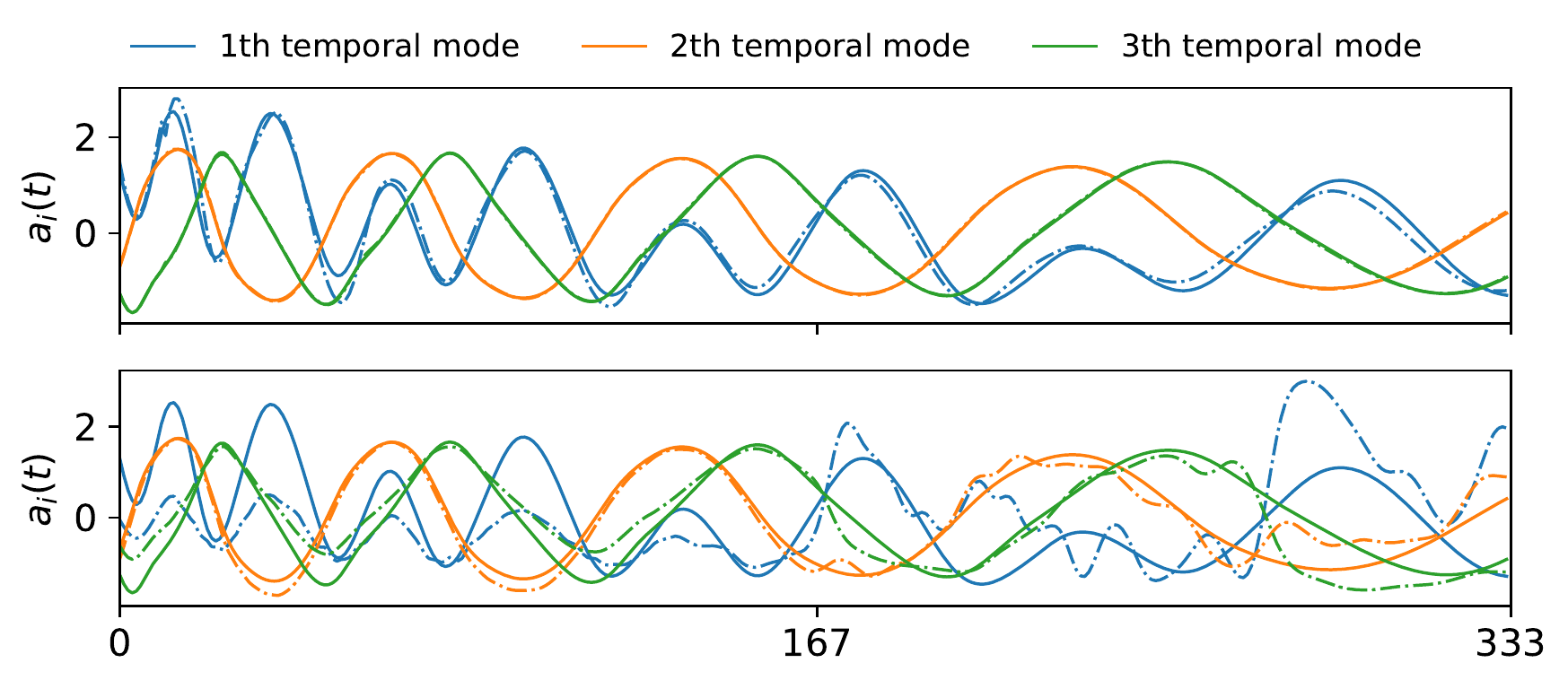}
    \caption{Predictions of the first three temporal amplitudes for the POD-LSTM models against the reference data when the rank of truncation is equal to 3 (top) and 50 (bottom). The data is normalized with mean and standard deviation of each amplitude. Solid lines represent reference data and dash-dotted lines represent the LSTM predictions.}
    \label{temporal}
\end{figure}

In the next step, we increase the size of the latent space to 50 and compare the performance of the models. Results are reported in \cref{tab:pod50} for the POD-LSTM model and the corresponding results for the autoencoder-LSTM model can be found in \cref{tmse}. For the oscillating airfoil, which is a periodic test case, consideration of a higher number of POD modes lead to a more accurate performance of the POD-LSTM method. However, for the non-periodic flow of the cylinder test 2, it leads to a sub-optimal performance where a lower value is obtained for $R^{2}$ of the predictions in comparison with the test with the rank truncation of 3. We observed that the LSTM network could obtain excellent predictions of the dynamics when 3 temporal modes are considered as the input. However, in the case where 50 temporal modes are predicted with the LSTM network, the predictions are not as accurate as before. It is due to the fact that higher-order POD modes are essentially higher-frequency modes; prediction of the temporal dynamics of these modes may be challenging for the LSTM network especially for non-periodic flows. In \cref{temporal}, we show the first three temporal modes and their predictions by the LSTM network when the rank of truncation is equal to 3 \cref{temporal} (top) and 50 \cref{temporal} (bottom). Each temporal mode is normalized with its mean and standard deviation for better representation. It can be observed that the predictions of these temporal modes, in the case of POD rank truncation of 50, are contaminated with higher-frequencies as the LSTM network also tries to predict higher-frequency temporal amplitudes. On the other hand, the autoencoder-LSTM model also provides excellent predictions in this case as can be seen in \cref{tmse}.

\begin{center}
    \begin{table}[hbtp]
    \caption{$R^{2}$ and MSE for the POD-LSTM model in the prediction and reconstruction of the original test and train data. The latent space dimension $n_g$ is equal to 50. Corresponding results for the autoencoder-LSTM model have been reported in \cref{tmse}.}\label{tab:pod50}
    \centering
    \begin{tabular}{@{\extracolsep{5pt}}cccc@{}}
    \hline
    Test cases & &  $R^{2}$     & MSE \\ 
      \hline
      \multirow{ 2}{*}{Cylinder 2}&Train &0.9995&$3.64\times 10^{-8}$ \\ 
      &Test&0.3771& $5.88\times10^{-6}$ \\
      \rule{0pt}{4ex} \multirow{ 2}{*}{Oscillating airfoil}&Train &0.9211& $1.45\times 10^{-3}$ \\ 
      &Test&0.9163& $1.54\times 10^{-3}$ \\
      \hline
    \end{tabular}
    \end{table}
\end{center}

It should be noted that the POD provides a set of orthogonal modes that are ordered based on their energy. In contrast, the autoencoder modes are correlated and they are not ordered; it is not applicable to identify the dominant autoencoder modes. Moreover, we observed that, for instance, it is not guaranteed that the three modes of an autoencoder network with three neurons at the bottleneck layer be completely similar to three of the modes of an autoencoder with more neurons at the bottleneck. In fact, the size of the bottleneck layer is {\it a priori} proposition with {\it a posteriori} validation. The feasibility of the selected size of the bottleneck layer can be evaluated only after the training of the network, and if the selected size is not appropriate, the training process should be conducted again. It is in contrast to the POD that the appropriate size of the latent space can be obtained after the computation of the POD. It is one of the drawbacks of using the autoencoder network for ROM that may increase the computation time. However, we showed that for the investigated test cases, even an autoencoder with only a few neurons at the bottleneck can obtain very good reconstructions of the velocity field. It may be possible to use probabilistic neural networks such as variational autoencoders (VAE) and generative adversarial networks (GAN) to only store minimal sufficient uncorrelated representations at the latent space.

\subsection{Prediction of Transient Dynamics}

In the next test setup, we investigate the applicability of the LSTM network in the prediction of the transient dynamics of fluid flows. To this end, we utilize a mean-field model for two-dimensional viscous flow past a circular cylinder at $Re = 100$ \cite{noack2003,loiseau2018} as a canonical test case for modal decomposition, reduced order modeling, and model identification \cite{bagheri2013,Rowley2017,brunton2016}. The temporal dynamics of the flow are provided based on two most energetic POD modes and an additional mode, called the
shift mode, which is included to capture the transient dynamics from the onset of vortex shedding to the periodic von K\'arm\'an street. The low-order model for the described system is given by:
\begin{subequations}
    \begin{align}
    \dot{x_{1}} & = \mu x_{1} - \omega x_{2} + A x_{1} x_{3} \\
    \dot{x_{2}} & = \omega x_{1} + \mu x_{2} + A x_{2} x_{3} \\
    \dot{x_{3}} & = - \lambda (x_{3} - x_{1}^{2} - x_{2}^{2}),
    \end{align}
\end{subequations}
where $\mu = 0.1$, $\omega = 1$, $A = -0.1$, and $\lambda = 10$. We solve this system from an initial condition on the slow manifold where $r = 0.05$, $\theta = 0$, $x_{1} = r~cos(\theta)$, $x_{2} = r~sin(\theta)$, and $x_{3} = x_{1}^{2} + x_{2}^{2}$ for 2000 time steps with $\Delta t = 0.1$. The first 340 time steps are used for training of the model. An LSTM network is constructed with one hidden layer containing 600 LSTM cells and 60 previous time steps are used for one step prediction. The model is trained with the same stopping criterion as before and then the whole time span of the reference data is predicted. The DMD method cannot predict the transient dynamics of this system as it is essentially a linear model.
\begin{figure*}[hbtp]
    \centering
    \includegraphics[width=0.8\textwidth]{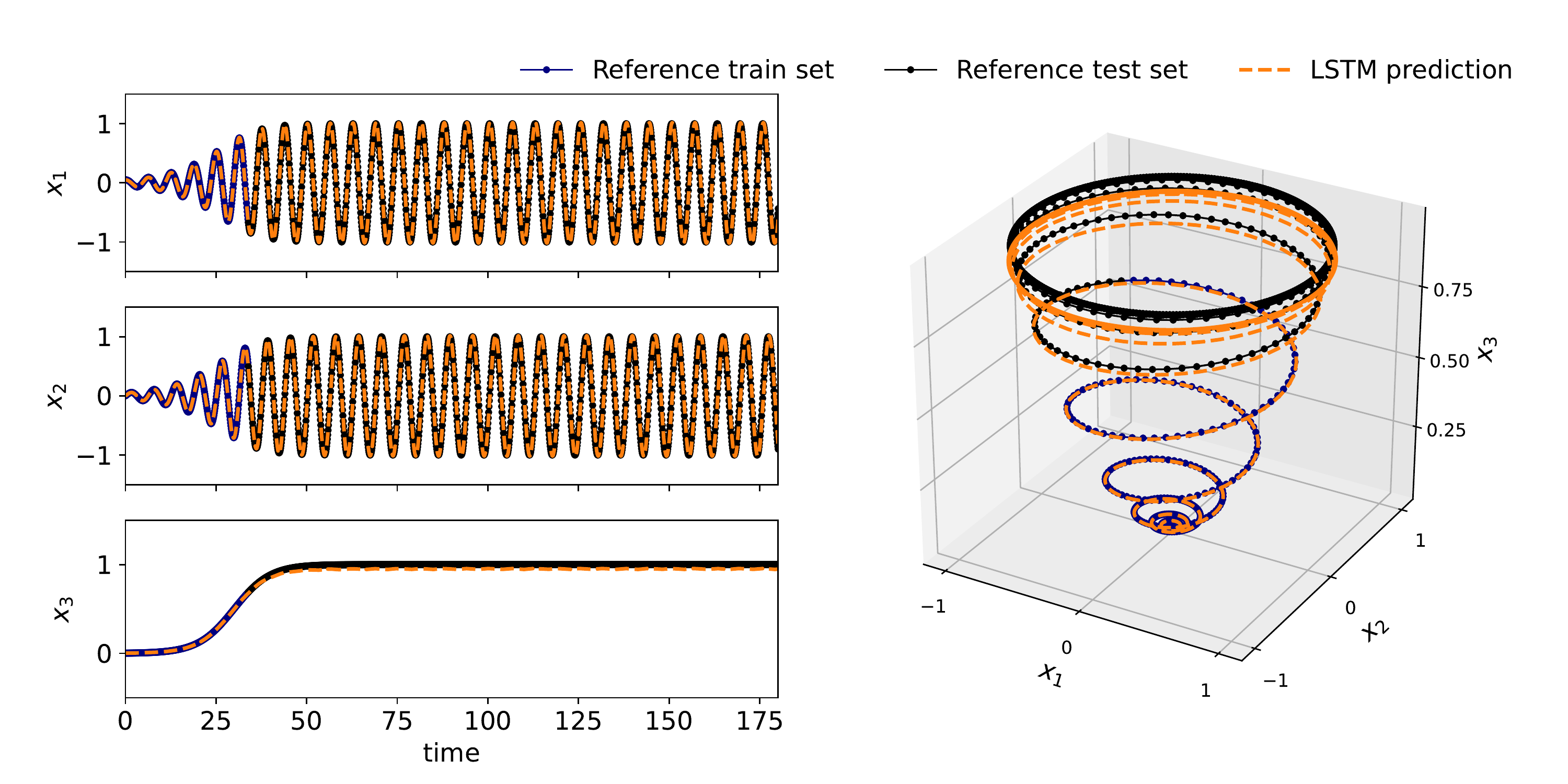}
    \caption{Transient dynamics of a low-order model of cylinder wake at $Re = 100$; prediction of the LSTM network against reference data.}
    \label{transient}
\end{figure*}
Results are depicted in \cref{transient}. It can be observed that although a small portion of the dynamics is used for training, the LSTM model can provide an excellent prediction of these nonlinear dynamics. These results suggest the possible applicability of the LSTM network for the prediction of the transient dynamics. However, since the investigated model here is a low-order model, the feasibility of the autoencoder-LSTM model for reduce order modeling of such fluid systems should be investigated through a more general model assessment.

\subsection{Computational Cost}
In this section, we discuss the computational efficiency of the proposed autoencoder-LSTM method and compare it with the DMD method. The main computational cost of the proposed ROM is the offline computation required for the training of the neural networks. The required time for both online and offline computations for each method are reported in \Cref{tab:time}. These costs are non-dimensional according to the time required for CFD simulations as $t_{\rm ROM} / t_{\rm CFD}$. Results are averaged over the three test cases of cylinder 1 and 2 and the oscillating airfoil. It can be observed that the autoencoder-LSTM method is two-order of magnitudes faster than CFD, which makes it viable for reduced order modeling of fluid flows. The DMD algorithm, however, is much faster than the autoencoder-LSTM, since it is trained in one shot, and it is five orders of magnitudes faster than CFD. Note that, we conducted our CFD simulations with RANS models, and in the case of direct numerical simulations (DNS) the computational costs of ROMs would be much lower than CFD.

\begin{center}
    \begin{table}[hbtp]
    \caption{Computational costs required for offline (training) and online (prediction) computations of the autoencoder-LSTM and the DMD methods. Times are non-dimensional according to the time required for CFD simulations as $t_{\rm ROM} / t_{\rm CFD}$.}\label{tab:time}
    \centering
    \resizebox{\columnwidth}{!}{%
    \begin{tabular}{@{\extracolsep{5pt}}cccc@{}}
    \hline
    Model & $n_g$ &  Offline (autoencoder \& LSTM) & Online \\ 
      \hline
    \multirow{ 2}{*}{autoencoder-LSTM} & 3  & 1.86$\times 10^{-2}$ \& 1.18$\times 10^{-2}$ & 0.19$\times 10^{-2}$ \\ 
    & 50 & 3.95$\times 10^{-2}$ \& 2.98$\times 10^{-2}$ & 0.19$\times 10^{-2}$ \\
    \multirow{ 2}{*}{DMD}& 3 & 7.05$\times 10^{-5}$ & 1.56$\times 10^{-5}$ \\ 
    & 50 & 9.19$\times 10^{-5}$ & 2.66$\times 10^{-5}$ \\
      \hline
    \end{tabular}
    }
    \end{table}
\end{center}

\section{Conclusion}
In this paper, a novel data-driven reduced order method based on the power of deep neural networks is presented with the aim of future estate estimation of the complex unsteady fluid flows. The proposed method is based on the power of the autoencoder neural network in dimensionality reduction and feature extraction and the power of the LSTM network in the prediction of the sequential data. Training and testing data are obtained from CFD simulations using finite volume method. Three test cases are investigated; a cylinder at the constant Reynolds number of 3900, a cylinder for which the Reynolds number is decreased with the time from 3355 to 676, and an oscillating airfoil. In the first step, the autoencoder network has been used as a nonlinear dimension reduction technique to project the high dimensional data onto the low dimensional subspace. In this way, the essential features of the flow feasible for accurate reconstruction of the velocity field are extracted at the bottleneck layer of the autoencoder. Then, sequences of the extracted features implemented as the input of an LSTM network with the aim of future state estimation. The output of the LSTM network is the velocity field at the next time step. Results are compared with the results of the well-known DMD method and a newly developed ROM based on the POD modes and the LSTM network. The performance of each method is assessed with the use of the coefficient of determination $R^{2}$ and MSE. Moreover, for the DMD method, the use of the autoencoder network for dimensionality reduction instead of SVD rank truncation is assessed. Results indicate the excellent potential of deep neural networks for data-driven reduced order modeling of complex unsteady flows. For all the cases, the autoencoder-LSTM network obtains the best results in the prediction of the velocity field in future time instances. Results show that the DMD and autoencoder-LSTM method can predict the flow evolution of the first test case accurately. However, for the second and third test cases, the autoencoder-LSTM method outperforms the DMD method in the prediction of the flow dynamics. Our results show excellent performance of the LSTM network in the prediction of the transient dynamics of a low-order model of cylinder wake at $Re = 100$. Moreover, we showed that the predictions of an autoencoder-LSTM model with three neurons at the bottleneck layer contain more than three POD basis, indicating excellent performance of the autoencoder network in dimension reduction.

\section*{Data Availability}
The data that support the findings of this study are available from the corresponding author upon reasonable request.
\bibliography{References}

\end{document}